\documentclass[prx,superscriptaddress,aps,amsmath,amssymb,amsfonts,floatfix,reprint,raggedbottom,longbibliography]{revtex4-2}

\usepackage{graphicx}
\graphicspath{{Figures/}}
\usepackage{balance,wrapfig}
\usepackage{dcolumn}
\usepackage{xcolor}
\usepackage{lipsum}
\usepackage{soul}
\usepackage{braket}
\usepackage{multirow}
\usepackage{array}
\usepackage{booktabs}
\usepackage[ruled,vlined,linesnumbered]{algorithm2e}
\usepackage{footnote}
\usepackage{titlesec}
\usepackage{lmodern}
\usepackage{times}
\usepackage[utf8]{inputenc}
\usepackage[subpreambles=true]{standalone}
\usepackage[para]{footmisc}
\usepackage[misc]{ifsym}
\usepackage{mathtools}
\usepackage[colorlinks=true,citecolor=blue,linkcolor=blue,urlcolor=blue]{hyperref}
\usepackage{xr-hyper}

\DeclareUnicodeCharacter{2009}{\,}

\makeatletter
\newcommand*{\addFileDependency}[1]{
  \typeout{(#1)}
  \@addtofilelist{#1}
  \IfFileExists{#1}{}{\typeout{No file #1.}}
}
\makeatother

\makeatletter 
\renewcommand{\fnum@figure}{\textbf{Fig.~\thefigure}}
\makeatother

\makeatletter
\def\bbordermatrix#1{\begingroup \m@th
  \@tempdima 4.75\p@
  \setbox\z@\vbox{%
    \def\cr{\crcr\noalign{\kern2\p@\global\let\cr\endline}}%
    \ialign{$##$\hfil\kern2\p@\kern\@tempdima&\thinspace\hfil$##$\hfil
      &&\quad\hfil$##$\hfil\crcr
      \omit\strut\hfil\crcr\noalign{\kern-\baselineskip}%
      #1\crcr\omit\strut\cr}}%
  \setbox\tw@\vbox{\unvcopy\z@\global\setbox\@ne\lastbox}%
  \setbox\tw@\hbox{\unhbox\@ne\unskip\global\setbox\@ne\lastbox}%
  \setbox\tw@\hbox{$\kern\wd\@ne\kern-\@tempdima\left[\kern-\wd\@ne
    \global\setbox\@ne\vbox{\box\@ne\kern2\p@}%
    \vcenter{\kern-\ht\@ne\unvbox\z@\kern-\baselineskip}\,\right]$}%
  \null\;\vbox{\kern\ht\@ne\box\tw@}\endgroup}
\makeatother

\newcolumntype{L}{>{\arraybackslash}m{3.9 cm}}
\newcolumntype{C}{>{\centering\arraybackslash}m{3.2 cm}}
\newcolumntype{G}{>{\centering\arraybackslash}m{1.41 cm}}

\emergencystretch=\maxdimen
\begin{document}

\title{Probabilistic Computers for Neural Quantum States}
\author{Shuvro Chowdhury}
\affiliation{Department of Electrical and Computer Engineering, University of California, Santa Barbara, Santa Barbara, CA 93106, USA}
\author{Jasper Pieterse}
\affiliation{Radboud University, Institute for Molecules and Materials,   Heyendaalseweg 135, Nijmegen, The Netherlands}
\author{Navid Anjum Aadit}
\affiliation{Department of Electrical and Computer Engineering, University of California, Santa Barbara, Santa Barbara, CA 93106, USA}
\author{Shaila Niazi}
\affiliation{Department of Electrical and Computer Engineering, University of California, Santa Barbara, Santa Barbara, CA 93106, USA}
\author{Johan H. Mentink}
\affiliation{Radboud University, Institute for Molecules and Materials, Heyendaalseweg 135, Nijmegen, The Netherlands}  
\author{Kerem Y. Camsari}
\affiliation{Department of Electrical and Computer Engineering, University of California, Santa Barbara, Santa Barbara, CA 93106, USA}
\date{\today}

\begin{abstract}
Neural quantum states efficiently represent many-body wavefunctions with neural networks, but the cost of Monte Carlo sampling limits their scaling to large system sizes. Here we address this challenge by combining sparse Boltzmann machine architectures with probabilistic computing hardware. We implement a probabilistic computer on field-programmable gate arrays (FPGAs) and use it as a fast sampler for energy-based neural quantum states. For the two-dimensional transverse-field Ising model at criticality, we obtain accurate ground-state energies for lattices up to 
80$\times$80 (6400 spins) using a custom multi-FPGA cluster.
Furthermore, we introduce a dual-sampling algorithm to train deep Boltzmann machines, replacing intractable marginalization with conditional sampling over auxiliary layers. This enables the training of sparse deep models and improves parameter efficiency relative to shallow networks. We further implement this algorithm on a single FPGA,  demonstrating the training of deep Boltzmann machines for systems as large as $30 \times 30$ (900 spins). Together, these results demonstrate that probabilistic hardware can overcome the sampling bottleneck in variational simulation of quantum many-body systems, opening a path to larger system sizes and deeper variational architectures. 
\end{abstract}


\maketitle

\twocolumngrid

\section{Introduction}
\label{sec:Intro}

\begin{figure*}[!ht]
    \centering
    \vspace{0pt}
    \includegraphics[width=0.95\textwidth,keepaspectratio]{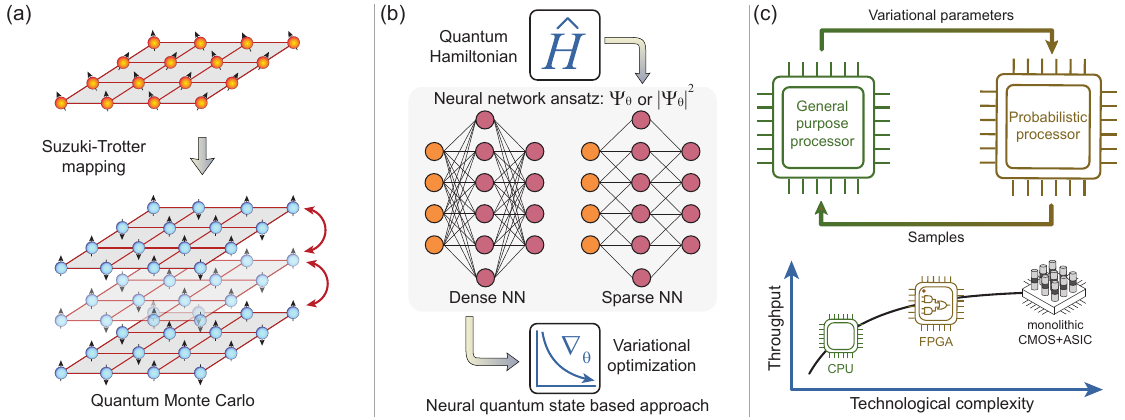}
    \caption{\footnotesize\textbf{Probabilistic computers for accelerating variational simulations of quantum systems.} (a) Conventional approaches such as path-integral quantum Monte Carlo (QMC) map quantum systems to classical statistical models in higher dimensions via the Suzuki-Trotter decomposition, but often suffer from sign problems or unfavorable scaling. (b) Neural Quantum States (NQS) represent the many-body wavefunction with neural networks ans\"{a}tze.  While many NQS architectures rely on dense or structured networks, sparse locally connected models are particularly well suited for mapping onto probabilistic hardware used in this work. Ground states are obtained through variational optimization of the quantum energy, which requires repeated Monte Carlo sampling from the neural network model. (c) Sampling is offloaded to a dedicated probabilistic processor built from p-bits. A general-purpose processor updates variational parameters while a p-bit array provides high-throughput samples. Throughput increases as implementations progress from CPU emulation to FPGA-based systems and, ultimately, to dedicated CMOS hardware.}
    \label{fig:fig1}
    \vspace{-5pt}
\end{figure*}

Many-body quantum simulation drives progress in condensed matter physics, quantum chemistry, and quantum information, yet accurate classical simulation remains challenging as system size grows. Established approaches such as quantum Monte Carlo (QMC) and tensor-network (TN) methods have achieved high precision in important regimes, but each faces fundamental limitations: sign problems for generic QMC and unfavorable entanglement scaling for tensor networks in two dimensions and near criticality \cite{troyer2005computational,SCHOLLWOCK201196,ORUS2014117}.

Neural Quantum States (NQS), introduced by Carleo and Troyer \cite{carleo_solving_2017}, offer a different approach  (Fig.~\ref{fig:fig1}a,b). By parameterizing many-body wavefunctions with neural networks, NQS can represent complex correlations in a compact and systematically improvable form \cite{LeRoux2008representational,schmitt2022quantum}. This framework has enabled progress in two-dimensional quantum systems beyond the reach of exact diagonalization \cite{Fabiani2021super,nomura2021dirac}.

Since their inception, NQS architectures have diversified. While early work relied on Restricted Boltzmann Machines (RBMs), subsequent studies adopted convolutional networks to exploit translational symmetry \cite{Choo2018symmetries}, recurrent neural networks \cite{hibat_2020_recurrent}, transformer-based models \cite{viteritti2023transformer,zhang2023transformer,zhao2021overcoming} and more recently foundation models \cite{rende2025foundation}, which significantly increased computational costs compared to RBMs. Despite this diversity, RBMs remain a canonical ansatz and continue to be used in studies of topological phases and fracton models \cite{Chen2025representing,Machaczek2025neural}.

Early NQS applications demonstrated simulations of systems with hundreds of spins, far beyond what is accessible with exact diagonalization. Scaling to larger system sizes remains an active area of research, with advances in software frameworks \cite{giammarco2019investigating,schmitt_quantum_2020,netket3_2022,jvmc_2022}, algorithms \cite{chen2024empowering,moss2025triangular,Moss2025leveraging}, and high-performance computing \cite{li2022bridging,Liang_2023_deep,xu2025largescale}, now reaching systems on the order of $10^3$ spins.

However, increasing system size by another order of magnitude remains challenging. For variational Monte Carlo (VMC), a widely used and expressive approach \cite{bortone2024impact}, scaling is constrained by the cost of Markov Chain Monte Carlo (MCMC) sampling required to estimate observables and stochastic parameter gradients. This bottleneck persists even for relatively simple architectures such as RBMs and becomes more severe for deeper or more structured models.

Probabilistic hardware provides a promising route to overcome this limitation. The joint energy of an RBM corresponds to an Ising Hamiltonian on a bipartite graph, allowing nonlocal correlations to be mediated through local pairwise interactions that map naturally onto probabilistic bits (p-bits) \cite{ACKLEY1985147,MEHTA20191}. P-bits are classical stochastic units that fluctuate between two logic states and can be implemented with massive parallelism and low-precision arithmetic \cite{camsari2017stochastic}. Recent theoretical work predicts large sampling speedups for NQS when using probabilistic hardware \cite{berns2025predicting}, and experimental demonstrations have shown orders-of-magnitude acceleration for frustrated Ising models using hardware-accelerated convolutional RBMs \cite{brahma2025hardware}. Prior p-bit architectures have established that local connectivity enables sparse Boltzmann machines to map efficiently onto hardware while avoiding communication bottlenecks \cite{aadit2022massively,niazi_training_2024}.

Here, we implement a probabilistic computer (p-computer) using field-programmable gate arrays (FPGAs) to accelerate sampling for neural quantum states (Fig.~\ref{fig:fig1}c). Rather than emulating stochasticity in software, we map sparse Boltzmann machines directly onto the FPGA fabric, exploiting spatial parallelism and low-precision arithmetic. Enforcing local connectivity enables efficient on-chip realization and avoids the routing congestion inherent to dense networks.

This work makes two complementary contributions. First, we map Further Restricted Boltzmann Machine (FRBM) $-$ a sparse variant of RBM $-$ onto probabilistic hardware. Using hardware-parallel sampling on a custom multi-FPGA cluster, we scale to lattices as large as $80 \times 80$ spins (6400) and obtain ground-state energies within chemical accuracy of variational benchmarks for the two-dimensional transverse-field Ising model at criticality.

Second, we introduce a dual-sampling algorithm for deep Boltzmann machines (DBMs) that replaces intractable marginalization over auxiliary variables with conditional sampling. This reformulation enables stable training of sparse deep architectures under strict locality constraints and improves parameter efficiency relative to shallow networks. We implement and validate this dual-sampling approach on a single FPGA, demonstrating DBM training for systems up to $30 \times 30$ spins (900).


Together, these advances address two barriers to scaling neural quantum states: probabilistic hardware alleviates the sampling bottleneck, while dual sampling makes sparse deep Boltzmann machines trainable for variational Monte Carlo.

\begin{figure*}[!t]
    \centering
    \includegraphics[width=0.95\textwidth,keepaspectratio]{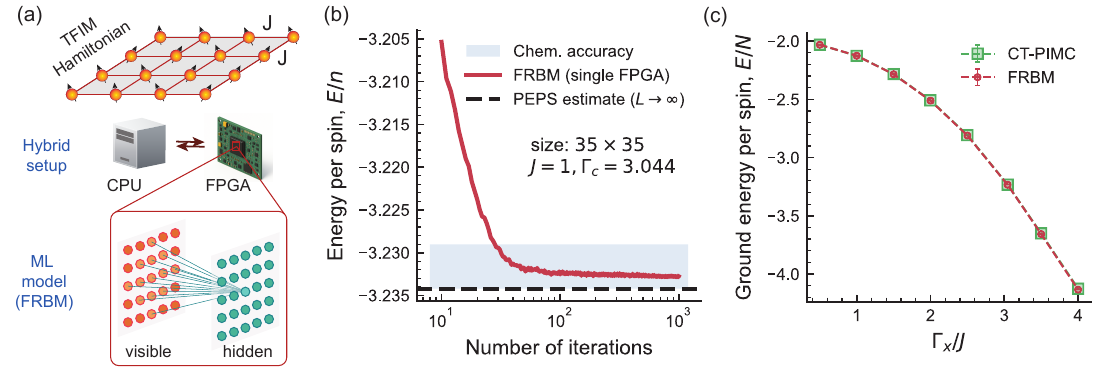}
    \caption{\footnotesize\textbf{Single-FPGA results for the 2D transverse-field Ising model.} (a) Problem and hybrid setup: a 2D square lattice with periodic boundaries, simulated using a CPU-FPGA platform where the FPGA samples from a Further Restricted Boltzmann Machine (FRBM) with local connectivity ($k=2$, corresponding to 13 neighbors). (b) Training convergence for a $35\times35$ lattice (1225 spins) at the critical field $\Gamma_c/J = 3.044$. The energy per spin reaches chemical accuracy (blue shaded band: relative error $|\Delta E/E_{\rm ref}| \leq 1.6\times 10^{-3}$, $E_{\rm ref}$ is a variational Projected Entangled Pair States (PEPS) benchmark estimate at the thermodynamic limit ($L\to \infty$) from Ref.~\cite{fedorovich2025finite}) within $\approx$100 iterations. (c) Ground-state energy per spin versus transverse field, interpolating between the ferromagnetic limit ($E/N \to -2$ as $\Gamma_x \to 0$) and the field polarised limit ($E/N \to -\Gamma_x$ as $\Gamma_x \to \infty$). For validation, we also compare against Continuous-Time Path Integral Monte Carlo simulation ($\beta=64$ and $10^5$ Monte Carlo sweeps) using the code provided in \cite{king2021scaling}. FRBM results are obtained from an average over $10^6$ samples of a single run after the training is completed. Error bars represent standard error via blocking (50 bins) and are smaller than the symbol size.}
\label{fig:fig2}
\vspace{-10pt}
\end{figure*}

\section{Model and Ansatz}
\label{sec:sec2}

We target the two-dimensional transverse-field Ising model (TFIM) on a square lattice with periodic boundary conditions, defined by the Hamiltonian \cite{sachdev2011quantum}:
\begin{align}
\hat{H} = -J \sum_{\langle i,j \rangle} \hat{\sigma}^z_i \hat{\sigma}^z_j - \Gamma_x \sum_i \hat{\sigma}^x_i
\end{align}
where $\langle i,j \rangle$ denotes unique nearest-neighbor bonds, $J$ is the coupling strength, and $\Gamma_x$ is the transverse field. We define the many-body basis states $S = \{s_1, s_2, \dots, s_N\}$ where $s_i \in \{+1, -1\}$ and parameterize the squared wavefunction amplitude $|\Psi_{\theta}(S)|^2$ through the Boltzmann distribution of a probabilistic network \cite{carleo_solving_2017}:
\begin{align}
|\Psi_{\theta}(S)|^2 = P_{\theta}(S) = \frac{1}{Z_{\theta}} \sum_{h}{e^{-E_{\theta}(S,h)}}
\end{align}
where $h=\{h_1,\dots,h_M\}$ with $h_j\in\{+1,-1\}$. This mapping relies on the Perron-Frobenius theorem, which guarantees that the ground state of stoquastic Hamiltonians like the TFIM can be chosen to have non-negative real amplitudes \cite{marshall1955anti}. The quantum probability distribution $|\Psi|^2$  then coincides with  the classical distribution generated by p-bits. As a result, the p-computer's stochastic fluctuations directly generate samples from $P_{\theta}(S)=|\Psi_{\theta}(S)|^2$ without computing amplitudes explicitly. Extending this approach to non-stoquastic systems with nontrivial sign structures would require complex parameters or phase networks \cite{bravyi2008complexity, szabo2020neural, Torlai2018neural}.

The variational ground-state energy of a neural quantum state (NQS) $\Psi_{\theta}(S)$ is evaluated as
\begin{equation}E = \frac{\langle \Psi_{\theta} | H | \Psi_{\theta} \rangle}{\langle \Psi_{\theta} | \Psi_{\theta} \rangle}= \mathbb{E}_{S \sim |\Psi_{\theta}|^2}\!\left[ E_{\mathrm{loc}}(S) \right]
\label{eq:var_energy}
\end{equation}
where the local energy is defined as
\begin{equation}
E_{\mathrm{loc}}(S)= \sum_{S'} \frac{\Psi_{\theta}(S')}{\Psi_{\theta}(S)}\langle S | H | S' \rangle
\end{equation}
Training the neural quantum state corresponds to minimizing this variational energy $E(\theta)$ with respect to the network parameters $\theta$. Expectation values are obtained by Monte Carlo sampling of configurations $S$ drawn from the probability distribution $|\Psi_{\theta}(S)|^2$. For the transverse-field Ising model, the off-diagonal term $-\Gamma_x \sum_i \sigma_i^x$ couples a configuration $S$ to configurations $S^{(i)}$ that differ by a single spin flip at site $i$. As a result, evaluating $E_{\mathrm{loc}}(S)$ requires computing wavefunction ratios of the form $\Psi_{\theta}(S^{(i)})/\Psi_{\theta}(S)$ for sampled configurations. These ratios also appear in estimators of physical observables such as $\langle \sigma_i^x \rangle$. For shallow architectures such as RBMs, these wavefunction ratios can be evaluated analytically. However, for deep Boltzmann machines, the wavefunction amplitude involves an intractable marginalization over hidden units, rendering direct evaluation of such ratios prohibitively expensive. To overcome this bottleneck, we introduce in Section~\ref{sec:arch_based} a dual sampling strategy that estimates wavefunction ratios via conditional sampling over hidden variables while preserving unbiased estimators of the local energy.

\section{Sparse Architecture and Hardware Implementation}

The physical implementation (Fig.~\ref{fig:fig2}a) uses a hybrid probabilistic-classical architecture \cite{niazi_training_2024,chowdhury2025pushing}: a Xilinx Alveo U250 FPGA serves as the probabilistic processor while a host CPU handles parameter optimization. The U250 scales to large quantum systems through the sparse architecture described below.

Central to this approach is the Further Restricted Boltzmann Machine (FRBM) \cite{deng2017machine} with strictly local connectivity. Standard all-to-all RBMs require $\mathcal{O}(N^2)$ connections, creating routing congestion on hardware. The FRBM instead defines the network energy $E_{\theta}(S, h)$ as:
\begin{align}
E_{\theta}(S, h) = -\sum_{i} a_i s_i - \sum_{j} b_j h_j - \sum_{\langle i,j \rangle_k} W_{ij} s_i h_j
\label{eq:FRBM_energy}
\end{align}
where $h = \{h_1, h_2, \dots, h_M\}$ represents the hidden spins, $a, b, W$ are the variational parameters, and $\langle i,j \rangle_k$ restricts connections to nodes within Euclidean distance $k$. This reduces wiring complexity to $\mathcal{O}(N)$ for fixed $k$. Throughout this work we use $k=2$, corresponding to 13 neighbors per spin.  This sparsity maps directly onto the FPGA while preserving analytical tractability for wavefunction amplitudes, local energies, and gradients \cite{carleo_solving_2017, nomura2017restricted}.

To maximize p-bit density on the Alveo U250, we use 10-bit fixed-point arithmetic (1 sign, 6 integer, 3 fractional bits) for weights and states, reducing logic utilization per neuron and enabling greater parallelism than floating-point implementations. For uncorrelated stochasticity, we integrate the xoshiro pseudo-random number generator \cite{blackman2021scrambled}, which provides sufficient statistical quality to avoid sampling bias while maintaining a compact footprint \cite{chowdhury_accelerated_2023}. These choices allow a large ensemble of p-bits on a single device without compromising the precision needed for accurate energy estimation. On the host CPU, the evaluation of local energies, gradients, and stochastic reconfiguration updates is performed in single-precision floating point (FP32), providing sufficient numerical accuracy while maintaining high throughput.

Parameter optimization uses the Stochastic Reconfiguration (SR) algorithm \cite{sorella1998green, carleo_solving_2017}. The variational optimization loop is split between the FPGA and the host CPU: given a set of parameters, the FPGA generates spin configurations which are transferred to the host CPU for parameter updates, and the process is repeated until the variational energy $E(\theta)$ in Eq.~(\ref{eq:var_energy}) converges. To avoid inverting the dense Fisher Information Matrix, we use an iterative Conjugate Gradient (CG) solver with matrix-free matrix-vector products \cite{neuscamman2012optimizing}, which scales efficiently with network size. Because the CG steps involve simple local operations, future implementations could integrate parameter optimization directly onto the FPGA.

\begin{figure*}[!ht]
    \centering
    \vspace{0pt}
     \includegraphics[width=0.98\textwidth,keepaspectratio]{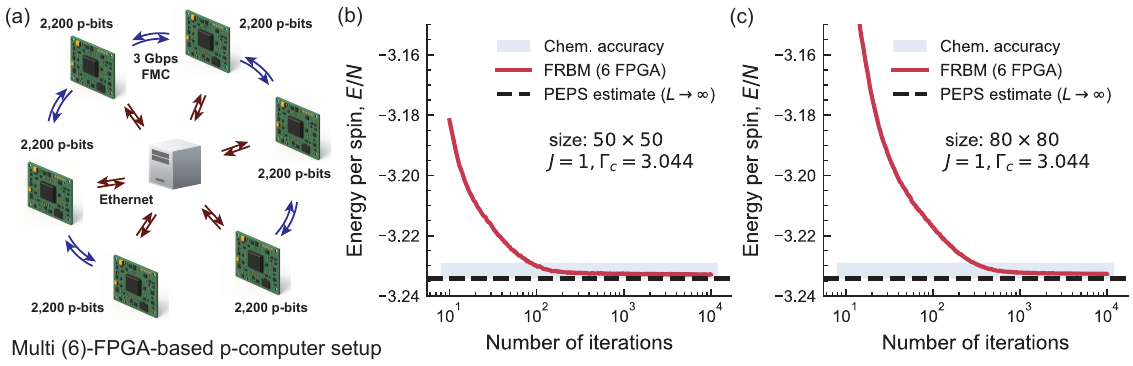}
  \caption{\footnotesize\textbf{Multi-FPGA results for large-scale neural quantum states.} (a) Six-FPGA probabilistic computer with boards connected via 3 Gbps FMC links and coordinated over Ethernet; each FPGA hosts up to 2200 p-bits. (b) Training convergence for a $50 \times 50$ lattice (2500 spins) at the critical field $\Gamma_c/J=3.044$. (c) Same for an $80 \times 80$ lattice (6400 spins), showing empirically that convergence within chemical accuracy (blue shaded region) is maintained as system size increases. PEPS benchmark estimate at the thermodynamic limit is taken from Ref.~\cite{fedorovich2025finite}.}
\label{fig:fig3}
\end{figure*}

\section{Validation on single-FPGA}

We validated the method on a $35 \times 35$ lattice (1225 spins, 2450 p-bits). As shown in Fig.~\ref{fig:fig2}b, the variational energy per spin converges to the benchmark ground-state value \cite{fedorovich2025finite}, reaching chemical accuracy (relative error $|\Delta E/E_{\rm ref}| \leq 1.6\times 10^{-3}$) within approximately 100 optimization iterations. Each iteration involves $10^4$ MCMC samples (see Supplementary Section~\ref{supp_sec:hyperparamters}). 

Profiling the training loop (Supplementary~Section~\ref{supp_sec:runtime_profiling}) shows that sampling consumes less than 5\% of total wall-clock time on the FPGA, despite executing $3\times10^6$ Monte Carlo sweeps per iteration. In contrast, a CPU baseline limited to $10^4$ sweeps per iteration already spends 20–30\% of runtime on sampling; matching the FPGA's throughput would be prohibitively slow. This reduction in sampling time aligns with theoretical predictions of large sampling speedups for stochastic Ising machines applied to NQS \cite{berns2025predicting}.

Crucially, the sampling time per sweep remains constant as system size increases, provided the FRBM fits within the resources of the probabilistic chip \cite{nikhar2024all, chowdhury2025pushing}. With sampling latency minimized, local energy computation and CG-based gradient accumulation dominate the remaining execution time. The sparse connectivity of our architecture opens a path to offloading local-energy calculations to hardware, an optimization we leave for future work. To further validate the solution, we swept the transverse field $\Gamma_x/J$ across the critical point \cite{blote2002cluster}. Fig.~\ref{fig:fig2}c shows the ground-state energy per spin interpolating smoothly between the ferromagnetic limit (low $\Gamma_x$) and the field-polarized limit (high $\Gamma_x$).  These results indicate that sparse connectivity, together with low-precision sampling, provides sufficient accuracy for large lattices.

\section{Large-Scale Systems via Multi-FPGA Clusters}
\label{sec:apt}

A single FPGA suffices for moderately sized lattices, but larger systems require distributing the computation. We developed a cluster of six interconnected FPGA boards (Supplementary Section~\ref{supp_sec:details_cluster}), partitioning the FRBM with the graph partitioning tool METIS to minimize the number of boundary p-bits communicated between devices.
Each FPGA updates its local p-bits using local fields and the most recently received boundary states, which are exchanged asynchronously and held constant between communication events.

Each FPGA hosts a subgraph containing approximately 850$-$2200 p-bits depending on the system size. METIS partitioning yields cut fractions of $8.6\%$ for $L=50$ and $5.6\%$ for $L=80$, so that most interactions remain on-chip. The FPGAs are connected in a linear topology via $3~\mathrm{Gbps}$ FMC links. Boundary p-bits are exchanged asynchronously over the FMC links while local p-bits update synchronously within each FPGA.

Strict global synchronization would limit the local p-bit update clock to $2.4~\mathrm{MHz}$ for $L=50$ and $1.2~\mathrm{MHz}$ for $L=80$. In practice, because boundary p-bits constitute a small fraction of the lattice, the system tolerates delayed boundary updates, allowing local p-bits to be overclocked to $15~\mathrm{MHz}$. A host computer loads weights and reads out states over Ethernet.

Using this architecture, we simulate the 2D TFIM at the critical point $\Gamma_c/J = 3.044$ for $50 \times 50$ (2500 spins) and $80 \times 80$ (6400 spins) lattices (Fig.~\ref{fig:fig3}b,c). These results show that p-bit-accelerated NQS reaches system sizes beyond those reported for CPU- and GPU-based implementations while maintaining accuracy within chemical precision of variational benchmarks. 

The convergence behavior remains comparable across system sizes: for both lattices, the variational energy density converges smoothly to chemical accuracy (blue shaded region). The accuracy on the $80\times80$ lattice confirms that the distributed architecture and asynchronous communication do not introduce artifacts that degrade the solution.

\section{Deep Neural Quantum States via Dual Sampling}
\label{sec:arch_based}

\SetKwComment{Comment}{\normalfont $\triangleright$ }{}
\SetKwInput{KwInput}{Input}                
\SetKwInput{KwOutput}{Output}              
\SetKwFor{ParFor}{for}{do in parallel}{end}
\SetKwFor{For}{for}{}{end}

\begin{algorithm}[!ht]
\DontPrintSemicolon
\caption{Dual Sampling Algorithm for Deep NQS}
\label{algo:dual_sampling_short}

\KwInput{Hamiltonian parameters, Learning rate $\eta$, Samples $N_s$ (outer), $N_c$ (inner)}
\KwOutput{Optimized parameters $\theta$ (weights and biases)}

\For{each optimization step $t = 1$ to  $N_{\rm iter}$}{
    Generate $N_s$ visible configurations $v$ by sampling the physical layer.\;
    
    \For{each visible configuration $v$}{
        Clamp visible units to fixed state $v$. \;
        Run $N_c$ Gibbs steps for auxiliary layers $(h, d)$.\;
        Accumulate conditional expectations for wavefunction ratios and gradients. \;
        Compute local energy $E_{\rm loc}(v)$ by combining single-spin-flip ratios.\;
    }
    Compute natural gradients $\nabla \theta$ using Stochastic Reconfiguration (SR).\;
    Update parameters: $\theta \leftarrow \theta - \eta(t) \nabla \theta$.\;
}

\KwRet Optimized $\theta$ and $E_{\mathrm{G}}^{(N_{\mathrm{iter}})}$.\;
\end{algorithm}

\begin{figure*}[!ht]
    \centering
    \vspace{0pt}
    \includegraphics[width=0.98\textwidth,keepaspectratio]{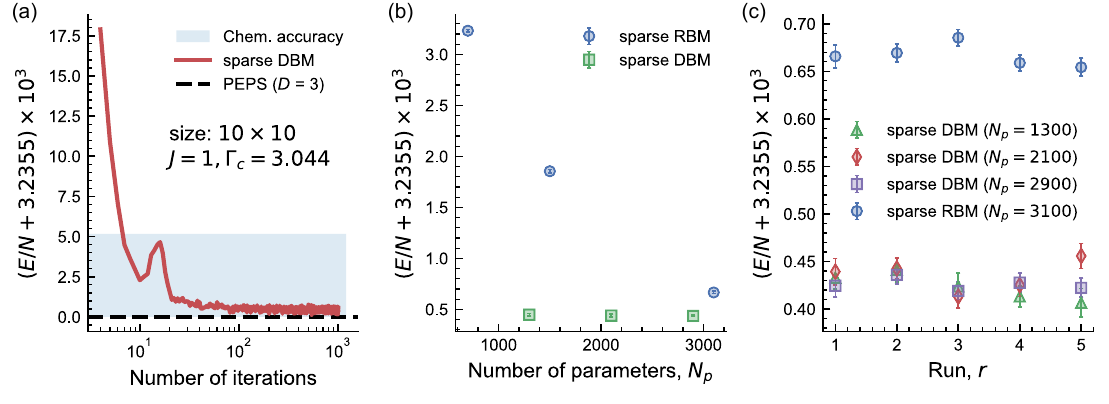}
    \caption{\footnotesize\textbf{Deep Boltzmann machine results for the 2D transverse-field Ising model.} (a) Convergence of the sparse DBM using dual sampling on a $10 \times 10$ lattice ($J = 1$, $\Gamma_c = 3.044$). Supplementary Section~\ref{supp_sec:hyperparamters} provides the hyperparameter choices. The variational energy converges toward a variational PEPS benchmark from Ref.~\cite{fedorovich2025finite}. A transient non-monotonic deviation appears early in training due to the relatively aggressive learning-rate and SR regularization parameters used in this demonstration. This effect is most visible in the small-system DBM case shown here, where stochastic fluctuations and sensitivity to initialization are larger; with more conservative optimization settings the trajectory becomes smoother. For the larger systems and shallow architectures shown elsewhere in the manuscript, the optimization trajectory is correspondingly smoother. (b) Final converged energy versus number of variational parameters $N_p$ for the sparse DBM (green squares) and sparse RBM (blue circles). Parameter count is varied by adjusting the connectivity radius $k$ (see Supplementary Section~\ref{supp_sec:parameter_counting}). The most compact DBM ($N_p \approx 1300$) achieves lower energies than RBMs with more than twice as many parameters. (c) Final energy across 5 independent training runs with random initializations. The DBM consistently converges to lower energies than the RBM. Error bars in (b) and (c) represent standard error via blocking (50 bins); the y-axis shows residual energy scaled by $10^3$ relative to the offset $-3.2355$.}
\label{fig:fig4}
\end{figure*}

While the shallow FRBM suffices for the TFIM, more complex phases with topological order or frustration require correlations beyond a finite local horizon. On hardware with strict locality constraints, shallow networks face a geometric ceiling that width alone cannot overcome. Deep Boltzmann Machines (DBMs) address this by stacking local layers to expand receptive fields, allowing global correlations to emerge while preserving hardware-compatible sparsity. Theory supports this: deep architectures can efficiently encode volume-law entanglement and states from polynomial-depth quantum circuits inaccessible to shallow RBMs \cite{gao2017efficient}. Constructive proofs further show that DBMs can exactly represent ground states where shallow networks fail \cite{carleo2018constructing}. Extending the FRBM energy in Eq.~(\ref{eq:FRBM_energy}), we define a sparse DBM energy by coupling the hidden layer $h$ to an additional deep layer $d$:

\begin{align}
E_{\theta}(S, h, d) &= -\sum_{i} a_i s_i - \sum_{j} b_j h_j - \sum_{\langle i,j \rangle_{k_1}} W^{(vh)}_{ij} s_i h_j \nonumber \\
&- \sum_{l} c_l d_l - \sum_{\langle j,l \rangle_{k_2}} h_j W^{(hd)}_{jl} d_l
\end{align}
where $W^{(hd)}$ connects local neighbors in layers $h$ and $d$.

This depth precludes the analytic marginalization that makes local energy and gradient calculations tractable in RBMs. Evaluating local energy and gradients in DBMs requires nested sampling over auxiliary variables~\cite{salakhutdinov2009deep}, which has limited the adoption of DBMs for neural quantum states~\cite{nomura_purifying_2021}.

We address this challenge by developing a dual sampling algorithm (Algorithm~\ref{algo:dual_sampling_short}) that reformulates variational Monte Carlo for deep neural quantum states in terms of conditional expectations. This strategy decouples sampling of physical spins from conditional sampling of auxiliary layers, while preserving an asymptotically unbiased estimator of the variational energy.

Conceptually, this mirrors contrastive-learning-style training in energy-based models, where clamped (data-conditioned) correlations are estimated by sampling hidden variables conditioned on fixed visible data~\cite{hinton2002training,hinton2006fast}. In the present setting, the ``data'' corresponds to a fixed visible spin configuration $S$, and auxiliary variables are sampled conditionally for each such configuration. This algorithm exploits the sparse structure of DBMs and aligns naturally with fast probabilistic samplers.  

Within this framework, the wavefunction amplitude ratios are expressed as conditional expectations over the auxiliary variables given a fixed visible configuration:
\begin{align}
\frac{\Psi(S^{(i)})}{\Psi(S)} = \sqrt{ \frac{P(S^{(i)})}{P(S)} } = \sqrt{ \mathbb{E}_{(h,d) \sim P(h,d|S)} \left[ e^{- \Delta E_i(S, h,d)} \right] }
\end{align}
where $\Delta E_i = E(S^{(i)},h,d)-E(S,h,d)$ is the change in the classical energy of the DBM upon flipping a visible spin in $S$ at site $i$, and the expectation is estimated via samples drawn from the conditional distribution $P(h,d|S)$. 

\begin{figure*}[t!]
    \centering
    \vspace{0pt}
    \includegraphics[width=0.98\textwidth,keepaspectratio]{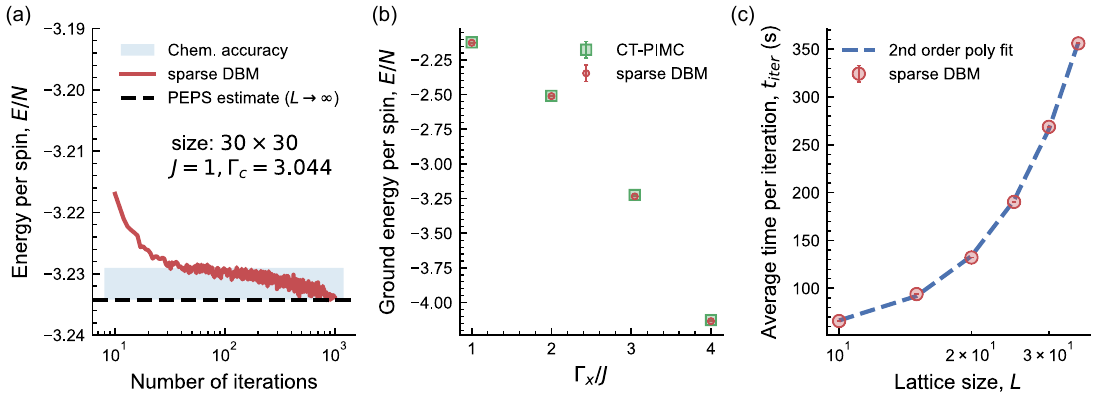}
    \caption{\footnotesize\textbf{Scalable training of Deep Boltzmann Machines (DBMs) on FPGA.} (a) Training convergence for a $30 \times 30$ lattice at the critical point $\Gamma_c/J \approx 3.044$. We train a sparse DBM with local connectivity radius $k=2$ on both layers (13 connections per spin), resulting in $N_p = 26,100$ variational parameters. The variational energy (red line) converges to within chemical accuracy (blue shaded region) of a variational PEPS benchmark estimate at the thermodynamic limit ($L\to \infty$) from Ref.~\cite{fedorovich2025finite}. (b) Ground-state energy density across the transverse-field phase transition for a $30 \times 30$ lattice, comparing the sparse DBM (red circles) against CT-PIMC (green squares, same as in Fig.~\ref{fig:fig2}). The DBM captures the phase transition with high precision. (c) Empirical algorithmic scaling of the dual-sampling estimator on a single NVIDIA V100 GPU. The average wall-clock time per optimization iteration (red circles, averaged over 500 iterations; error bars indicate 95\% confidence intervals) is shown as a function of the linear lattice dimension $L$. To strictly isolate algorithmic scaling from hardware parallelization effects, we use a single sequential MCMC chain rather than parallel independent chains. Under fixed sparsity and sampling budgets ($N_s, N_c$), the iteration time scales approximately as $t_{\mathrm{iter}} \propto L^2$ (dashed quadratic fit), corresponding to linear dependence on the total number of spins $N=L^2$. This behavior is consistent with the simplified cost model for conventional processors in Supplementary Section G.}
\label{fig:fig5}
\end{figure*}

Conditioning on the visible configuration substantially reduces variance by restricting sampling to the auxiliary variables. Reusing each of the $N_c$ conditional samples to evaluate all single-spin-flip ratios further reduces computational cost by avoiding resampling for each spin flip. This construction yields an asymptotically unbiased estimator of the wavefunction ratio that converges to the exact value in the infinite-sample limit (Supplementary Section~\ref{supp_sec:proof_dualSampling}). The nonlinear square-root mapping from probabilities to amplitudes introduces a residual bias at finite $N_c$. To mitigate this, we apply a second-order Taylor correction to the spin-flip probability estimates, as detailed in Supplementary Algorithm~\ref{algo:fullDualSampling}. 

We validate the approach on a $10 \times 10$ lattice,  using $10^3$ inner-loop samples to estimate local energy and  parameter derivatives. At this size,   optimization results can be directly compared against reliable benchmarks. As shown in Fig.~\ref{fig:fig4}a, the variational energy converges within chemical accuracy (blue shaded region) relative to a recent PEPS benchmark \cite{fedorovich2025finite}. For the system studied here, this empirical convergence indicates that $10^3$ conditional samples in the inner-loop suffice to maintain the correct balance between diagonal and off-diagonal contributions to the local energy, keeping stochastic reconfiguration updates well-conditioned. Supplementary Section~\ref{supp_sec:complexity} provides a computational complexity analysis of the algorithm.

Beyond convergence and accuracy, Fig.~\ref{fig:fig4}b compares representational efficiency between shallow and deep architectures. We swept the number of variational parameters $N_p$ for both the single-layer sparse RBM (blue circles) and the sparse DBM (green squares). While both architectures reach chemical accuracy, the sparse DBM achieves lower variational energies with less than half the parameters ($N_p \approx 1300$ versus $N_p \approx 3100$), consistent with theoretical results showing that depth provides more efficient representations \cite{gao2017efficient}. The improvement is modest here, likely because the TFIM has relatively simple entanglement structure; deeper gains may emerge for Hamiltonians with more complex correlations.

Finally, Fig.~\ref{fig:fig4}c shows the final energy across five independent training runs with random initializations. The sparse DBM consistently converges to lower energies, whereas the sparse RBM settles into higher-energy local minima. This reproducibility suggests that the additional layers help mediate interactions between distant lattice regions while preserving hardware-compatible sparsity. More generally, the dual-sampling framework is not restricted to RBM or layered DBM architectures, but applies to arbitrary Boltzmann-machine graphs, including networks with unstructured, interlayer, and hierarchical connectivity. This generality enables systematic exploration of expressive graph topologies for neural quantum states beyond the constraints of analytically tractable models.

We now evaluate the \emph{algorithmic} scaling of dual sampling under fixed sparsity and sampling budgets, independent of specialized sampling hardware. We use a conventional GPU to make sampling costs explicit and to measure how the dual-sampling estimator itself scales with system size on conventional processors.

This choice is conservative. Sparse, locally connected graphs are difficult to execute efficiently on GPUs due to irregular memory access and limited data reuse, unlike the dense linear algebra workloads for which GPUs are optimized~\cite{bell2009implementing, williams2009optimization, chen2018gunrock}. Stable convergence and predictable scaling under these conditions therefore, provide a stringent test of the dual-sampling estimator. 

Sparsity is also a design choice that reduces parameter count, arithmetic operations, and memory traffic per iteration, and aligns naturally with the probabilistic hardware architecture~\cite{horowitz2014computing, gale2019state}. The results in Fig.~\ref{fig:fig5} quantify this scaling leveraging the FPGA implementation, which reduces the sampling latency for large systems.

To demonstrate the scalability of the dual sampling algorithm, we trained a sparse DBM on a $30 \times 30$ lattice using an Xilinx Alveo U250 accelerator card.  Weights and biases with fixed-point precision of 10 bits (1 sign bit, 4 integer bits, and 5 fraction bits) are used for this experiment.  As shown in Fig.~\ref{fig:fig5}a, the variational energy converges within chemical accuracy of the benchmark ground-state energy at the critical point ($\Gamma_c/J \approx 3.044$). This stable convergence reflects the variance reduction achieved by conditioning the inner-loop sampling on fixed visible configurations.

The accuracy of the trained DBM extends across the phase diagram. Fig.~\ref{fig:fig5}b compares the DBM ground-state energy density against CT-PIMC benchmarks, showing agreement throughout both the ferromagnetic and field-polarized phases. 

Fig.~\ref{fig:fig5}c is an algorithmic check performed on a single NVIDIA V100 GPU that plots the average wall-clock time per iteration as a function of the linear lattice dimension $L$, revealing a quadratic dependence $t_{\mathrm{iter}}$\,$ \propto $\,$L^2$ under fixed sparsity and sampling budgets. Since the total number of spins scales as $N$\,$=$\,$L^2$, this behavior is consistent with the simplified cost analysis presented in Supplementary Section~\ref{supp_sec:complexity}. Probabilistic hardware can remove the explicit system-size dependence from sampling under the on-chip mapping assumption. However, the overall training cost remains linear in $N$ (quadratic in $L$) due to stochastic reconfiguration updates performed on the host processor. Further reducing the iteration time would require accelerating these optimization routines as well, for example through dedicated hardware support, which we leave for future work. 

\section{Conclusion}

We have demonstrated that probabilistic computers built from p-bits can extend neural quantum state simulations to system sizes beyond current software implementations. Mapping sparse Boltzmann machines onto a multi-FPGA cluster, we obtained ground-state energies within chemical accuracy for the 2D transverse-field Ising model at criticality, reaching 6400 spins. We also introduced dual sampling, an algorithm that replaces intractable marginalization in deep Boltzmann machines with conditional sampling, enabling the training of deep architectures with improved parameter efficiency relative to shallow networks.
These results suggest that spatial locality and asynchronous parallelism are the key computational resources for scalable variational Monte Carlo. The sparse connectivity that enables efficient hardware mapping also makes local energy evaluation embarrassingly parallel: each wavefunction ratio depends only on a bounded neighborhood, opening a path to computing $E_{\mathrm{loc}}$ entirely on-chip and eliminating the remaining data-transfer bottleneck. Similarly, the dual-sampling algorithm is naturally suited to hardware acceleration, since its inner loop consists entirely of conditional Gibbs sampling over sparse graphs.

As p-bit architectures mature from FPGA prototypes to dedicated CMOS circuits, these optimizations become increasingly attractive. Application-specific hardware could integrate sampling, local energy evaluation, and gradient accumulation on a single die, reducing both latency and energy consumption by orders of magnitude. Such a platform would make variational Monte Carlo practical for quantum systems far larger than those accessible today, positioning probabilistic computing as a scalable route to classically simulating quantum matter.

\appendix
\section*{Methods}
\label{sec:methods}

\subsection*{p-computing overview}
\label{subsec:p-computing_overview}
p-computing is based on a network of p-bits $\{\sigma_i\}$, each fluctuating between two  logical states ($\sigma_i \in \{-1, +1\}$). The interactions between p-bits govern their state updates according to the following equations \cite{camsari2017stochastic}:
\begin{align}
I_i = \sum_{j}W_{ij}\sigma_j+b_i \label{eq:synapse}\\
\sigma_i = \text{sgn}\left(\tanh(\beta I_i)-r_{[-1,1]}\right)
\label{eq:neuron}
\end{align}
where $W$, $b$, and $\beta$ represent the interconnection matrix, bias vector, and inverse temperature, respectively. $r_{[-1,1]}$ is a uniformly distributed random variable in the interval $[-1, 1]$. Together, equations (\ref{eq:synapse}) and (\ref{eq:neuron}) drive the network toward the Boltzmann distribution:
\begin{align}
p(\{\sigma_i\}) = \cfrac{1}{Z}\exp{\left(-\beta E(\{\sigma_i\})\right)}\label{eq:boltz_prob}\\
E(\{\sigma_i\}) =  -\sum_{i<j}{W_{ij}\sigma_i\sigma_j}-\sum_{i}b_i\sigma_i
\label{eq:ising_energy}
\end{align}
where \(p(\{\sigma_i\})\) is the probability and \(E(\{\sigma_i\})\) is the energy of the state \(\{\sigma_i\}\), and \(Z\) denotes the partition function.

\subsection{FPGA details}
We mapped the physics‑inspired, massively parallel \textit{p}-computer architecture of Ref.~\cite{aadit2022massively} onto a Xilinx Alveo U250  accelerator card using graph coloring to maximize parallelism on the sparse instances. All arithmetic is fixed‑point and uses s\(\{6\}\{3\}\) precision (1 sign, 6 integer, 3 fractional bits).  Custom RTLs were developed to implement the algorithm based on the p-computing architecture and synthesized, placed and routed with Xilinx Vivado/Vitis tool chain. Weights and biases are converted to this fixed-point precision in MATLAB before being sent to the FPGA.

\section*{DATA AVAILABILITY}
The data that support the findings of this study are available from the corresponding author upon reasonable request.

\section*{CODE AVAILABILITY}
The codes that support the findings of this study are available from the corresponding author upon reasonable request.

\section*{Acknowledgments}
SC, NAA and KYC acknowledge support from the Office of Naval Research (ONR) Young Investigator Program grant, the National Science Foundation (NSF) CAREER Award under grant number CCF 2106260, the Army Research Laboratory under grant number W911NF-24-1-0228, the Semiconductor Research Corporation (SRC) grant, and the ONR-MURI grant N000142312708. Use was made of computational facilities purchased with funds from the National Science Foundation (CNS-1725797) and administered by the Center for Scientific Computing (CSC). The CSC is supported by the California NanoSystems Institute and the Materials Research Science and Engineering Center (MRSEC; NSF DMR 2308708) at UC Santa Barbara. JP acknowledges support from Quantum Delta NL and ELLIS unit Nijmegen. JHM acknowledges funding from the VIDI project no. 223.157 (CHASEMAG) and KIC project no. 22016 which are (partly) financed by the Dutch Research Council (NWO), as well as support from the European Union Horizon 2020 and innovation program under the European Research Council ERC Grant Agreement No. 856538 (3D-MAGiC) and the Horizon Europe project no. 101070290 (NIMFEIA). 

\section*{Author Contributions}
SC, JHM and KYC conceived the study. SC and JP performed different parts of the simulations. NAA provided  support for the multi-FPGA implementations. SN implemented the dual sampling algorithm on FPGA. SC and JP wrote the initial draft of the manuscript with inputs from JHM and KYC. All authors contributed to improving the draft and participated in designing the experiments, analyzing the results, and editing the manuscript.  

\section*{Competing Interests}
The authors declare no competing interests.

\bibliographystyle{unsrtnat}

\begin{thebibliography}{0}
\providecommand{\natexlab}[1]{#1}
\providecommand{\url}[1]{\texttt{#1}}
\expandafter\ifx\csname urlstyle\endcsname\relax
  \providecommand{\doi}[1]{doi: #1}\else
  \providecommand{\doi}{doi: \begingroup \urlstyle{rm}\Url}\fi

\end{thebibliography}


\begin{thebibliography}{62}
\providecommand{\natexlab}[1]{#1}
\providecommand{\url}[1]{\texttt{#1}}
\expandafter\ifx\csname urlstyle\endcsname\relax
  \providecommand{\doi}[1]{doi: #1}\else
  \providecommand{\doi}{doi: \begingroup \urlstyle{rm}\Url}\fi

\bibitem[Troyer and Wiese(2005)]{troyer2005computational}
Matthias Troyer and Uwe-Jens Wiese.
\newblock {Computational Complexity and Fundamental Limitations to Fermionic Quantum Monte Carlo Simulations}.
\newblock \emph{Physical review letters}, 94\penalty0 (17):\penalty0 170201, 2005.

\bibitem[Schollwöck(2011)]{SCHOLLWOCK201196}
Ulrich Schollwöck.
\newblock The density-matrix renormalization group in the age of matrix product states.
\newblock \emph{Annals of Physics}, 326\penalty0 (1):\penalty0 96--192, 2011.

\bibitem[Orús(2014)]{ORUS2014117}
Román Orús.
\newblock A practical introduction to tensor networks: Matrix product states and projected entangled pair states.
\newblock \emph{Annals of Physics}, 349:\penalty0 117--158, 2014.

\bibitem[Carleo and Troyer(2017)]{carleo_solving_2017}
Giuseppe Carleo and Matthias Troyer.
\newblock Solving the quantum many-body problem with artificial neural networks.
\newblock \emph{Science}, 355\penalty0 (6325):\penalty0 602--606, February 2017.

\bibitem[Le~Roux and Bengio(2008)]{LeRoux2008representational}
Nicolas Le~Roux and Yoshua Bengio.
\newblock Representational power of restricted boltzmann machines and deep belief networks.
\newblock \emph{Neural Computation}, 20\penalty0 (6):\penalty0 1631--1649, 2008.

\bibitem[Schmitt et~al.(2022)Schmitt, Rams, Dziarmaga, Heyl, and Zurek]{schmitt2022quantum}
Markus Schmitt, Marek~M. Rams, Jacek Dziarmaga, Markus Heyl, and Wojciech~H. Zurek.
\newblock Quantum phase transition dynamics in the two-dimensional transverse-field ising model.
\newblock \emph{Science Advances}, 8\penalty0 (37):\penalty0 eabl6850, 2022.

\bibitem[Fabiani et~al.(2021)Fabiani, Bouman, and Mentink]{Fabiani2021super}
G.~Fabiani, M.~D. Bouman, and J.~H. Mentink.
\newblock Supermagnonic propagation in two-dimensional antiferromagnets.
\newblock \emph{Phys. Rev. Lett.}, 127:\penalty0 097202, Aug 2021.

\bibitem[Nomura and Imada(2021)]{nomura2021dirac}
Yusuke Nomura and Masatoshi Imada.
\newblock Dirac-type nodal spin liquid revealed by refined quantum many-body solver using neural-network wave function, correlation ratio, and level spectroscopy.
\newblock \emph{Phys. Rev. X}, 11:\penalty0 031034, Aug 2021.

\bibitem[Choo et~al.(2018)Choo, Carleo, Regnault, and Neupert]{Choo2018symmetries}
Kenny Choo, Giuseppe Carleo, Nicolas Regnault, and Titus Neupert.
\newblock Symmetries and many-body excitations with neural-network quantum states.
\newblock \emph{Phys. Rev. Lett.}, 121:\penalty0 167204, Oct 2018.

\bibitem[Hibat-Allah et~al.(2020)Hibat-Allah, Ganahl, Hayward, Melko, and Carrasquilla]{hibat_2020_recurrent}
Mohamed Hibat-Allah, Martin Ganahl, Lauren~E. Hayward, Roger~G. Melko, and Juan Carrasquilla.
\newblock Recurrent neural network wave functions.
\newblock \emph{Phys. Rev. Res.}, 2:\penalty0 023358, Jun 2020.

\bibitem[Viteritti et~al.(2023)Viteritti, Rende, and Becca]{viteritti2023transformer}
Luciano~Loris Viteritti, Riccardo Rende, and Federico Becca.
\newblock Transformer variational wave functions for frustrated quantum spin systems.
\newblock \emph{Phys. Rev. Lett.}, 130:\penalty0 236401, Jun 2023.

\bibitem[Zhang and Di~Ventra(2023)]{zhang2023transformer}
Yuan-Hang Zhang and Massimiliano Di~Ventra.
\newblock Transformer quantum state: A multipurpose model for quantum many-body problems.
\newblock \emph{Phys. Rev. B}, 107:\penalty0 075147, Feb 2023.

\bibitem[Zhao et~al.(2021)Zhao, De, Chen, Stokes, and Veerapaneni]{zhao2021overcoming}
Tianchen Zhao, Saibal De, Brian Chen, James Stokes, and Shravan Veerapaneni.
\newblock Overcoming barriers to scalability in variational quantum monte carlo.
\newblock In \emph{Proceedings of the International Conference for High Performance Computing, Networking, Storage and Analysis}, SC '21, New York, NY, USA, 2021. Association for Computing Machinery.

\bibitem[Rende et~al.(2025)Rende, Viteritti, Becca, Scardicchio, Laio, and Carleo]{rende2025foundation}
Riccardo Rende, Luciano~Loris Viteritti, Federico Becca, Antonello Scardicchio, Alessandro Laio, and Giuseppe Carleo.
\newblock Foundation neural-networks quantum states as a unified ansatz for multiple hamiltonians.
\newblock \emph{Nature Communications}, 16\penalty0 (1):\penalty0 7213, Aug 2025.

\bibitem[Chen et~al.(2025)Chen, Yan, and Cui]{Chen2025representing}
Penghua Chen, Bowen Yan, and Shawn~X. Cui.
\newblock Representing arbitrary ground states of the toric code by a restricted boltzmann machine.
\newblock \emph{Phys. Rev. B}, 111:\penalty0 045101, Jan 2025.

\bibitem[Machaczek et~al.(2025)Machaczek, Pollet, and Liu]{Machaczek2025neural}
Marc Machaczek, Lode Pollet, and Ke~Liu.
\newblock {Neural quantum state study of fracton models}.
\newblock \emph{SciPost Phys.}, 18:\penalty0 112, 2025.

\bibitem[Fabiani and Mentink(2019)]{giammarco2019investigating}
Giammarco Fabiani and Johan~H. Mentink.
\newblock {Investigating ultrafast quantum magnetism with machine learning}.
\newblock \emph{SciPost Phys.}, 7:\penalty0 004, 2019.

\bibitem[Schmitt and Heyl(2020)]{schmitt_quantum_2020}
Markus Schmitt and Markus Heyl.
\newblock Quantum {Many}-{Body} {Dynamics} in {Two} {Dimensions} with {Artificial} {Neural} {Networks}.
\newblock \emph{Physical Review Letters}, 125\penalty0 (10):\penalty0 100503, September 2020.
\newblock Publisher: American Physical Society.

\bibitem[Vicentini et~al.(2022)Vicentini, Hofmann, Szabó, Wu, Roth, Giuliani, Pescia, Nys, Vargas-Calderón, Astrakhantsev, and Carleo]{netket3_2022}
Filippo Vicentini, Damian Hofmann, Attila Szabó, Dian Wu, Christopher Roth, Clemens Giuliani, Gabriel Pescia, Jannes Nys, Vladimir Vargas-Calderón, Nikita Astrakhantsev, and Giuseppe Carleo.
\newblock {NetKet 3: Machine Learning Toolbox for Many-Body Quantum Systems}.
\newblock \emph{SciPost Phys. Codebases}, page~7, 2022.

\bibitem[Schmitt and Reh(2022)]{jvmc_2022}
Markus Schmitt and Moritz Reh.
\newblock {jVMC: Versatile and performant variational Monte Carlo leveraging automated differentiation and GPU acceleration}.
\newblock \emph{SciPost Phys. Codebases}, page~2, 2022.

\bibitem[Chen and Heyl(2024)]{chen2024empowering}
Ao~Chen and Markus Heyl.
\newblock Empowering deep neural quantum states through efficient optimization.
\newblock \emph{Nature Physics}, 20\penalty0 (9):\penalty0 1476--1481, Sep 2024.
\newblock ISSN 1745-2481.

\bibitem[Moss et~al.(2025{\natexlab{a}})Moss, Wiersema, Hibat-Allah, Carrasquilla, and Melko]{moss2025triangular}
M.~Schuyler Moss, Roeland Wiersema, Mohamed Hibat-Allah, Juan Carrasquilla, and Roger~G. Melko.
\newblock Leveraging recurrence in neural network wavefunctions for large-scale simulations of heisenberg antiferromagnets on the triangular lattice.
\newblock \emph{Phys. Rev. B}, 112:\penalty0 134449, Oct 2025{\natexlab{a}}.

\bibitem[Moss et~al.(2025{\natexlab{b}})Moss, Wiersema, Hibat-Allah, Carrasquilla, and Melko]{Moss2025leveraging}
M.~Schuyler Moss, Roeland Wiersema, Mohamed Hibat-Allah, Juan Carrasquilla, and Roger~G. Melko.
\newblock Leveraging recurrence in neural network wavefunctions for large-scale simulations of heisenberg antiferromagnets on the square lattice.
\newblock \emph{Phys. Rev. B}, 112:\penalty0 134450, Oct 2025{\natexlab{b}}.

\bibitem[Li et~al.(2022)Li, Chen, Xiao, Wang, Jiang, Zhao, Lin, An, Liang, and He]{li2022bridging}
Mingfan Li, Junshi Chen, Qian Xiao, Fei Wang, Qingcai Jiang, Xuncheng Zhao, Rongfen Lin, Hong An, Xiao Liang, and Lixin He.
\newblock Bridging the gap between deep learning and frustrated quantum spin system for extreme-scale simulations on new generation of sunway supercomputer.
\newblock \emph{IEEE Transactions on Parallel and Distributed Systems}, 33\penalty0 (11):\penalty0 2846--2859, 2022.

\bibitem[Liang et~al.(2023)Liang, Li, Xiao, Chen, Yang, An, and He]{Liang_2023_deep}
Xiao Liang, Mingfan Li, Qian Xiao, Junshi Chen, Chao Yang, Hong An, and Lixin He.
\newblock Deep learning representations for quantum many-body systems on heterogeneous hardware.
\newblock \emph{Machine Learning: Science and Technology}, 4\penalty0 (1):\penalty0 015035, mar 2023.

\bibitem[Xu et~al.(2025)Xu, Wu, Li, and Jia]{xu2025largescale}
Hongtao Xu, Zibo Wu, Mingzhen Li, and Weile Jia.
\newblock Large-scale neural network quantum states for ab initio quantum chemistry simulations on fugaku, 2025.

\bibitem[Bortone et~al.(2024)Bortone, Rath, and Booth]{bortone2024impact}
Massimo Bortone, Yannic Rath, and George~H. Booth.
\newblock Impact of conditional modelling for a universal autoregressive quantum state.
\newblock \emph{Quantum}, 8:\penalty0 1245, Feb 2024.

\bibitem[Ackley et~al.(1985)Ackley, Hinton, and Sejnowski]{ACKLEY1985147}
David~H. Ackley, Geoffrey~E. Hinton, and Terrence~J. Sejnowski.
\newblock A learning algorithm for boltzmann machines.
\newblock \emph{Cognitive Science}, 9\penalty0 (1):\penalty0 147--169, 1985.

\bibitem[Mehta et~al.(2019)Mehta, Bukov, Wang, Day, Richardson, Fisher, and Schwab]{MEHTA20191}
Pankaj Mehta, Marin Bukov, Ching-Hao Wang, Alexandre~G.R. Day, Clint Richardson, Charles~K. Fisher, and David~J. Schwab.
\newblock A high-bias, low-variance introduction to machine learning for physicists.
\newblock \emph{Physics Reports}, 810:\penalty0 1--124, 2019.
\newblock A high-bias, low-variance introduction to Machine Learning for physicists.

\bibitem[Camsari et~al.(2017)Camsari, Faria, Sutton, and Datta]{camsari2017stochastic}
Kerem~Yunus Camsari, Rafatul Faria, Brian~M Sutton, and Supriyo Datta.
\newblock {Stochastic p-bits for invertible logic}.
\newblock \emph{Physical Review X}, 7\penalty0 (3):\penalty0 031014, 2017.

\bibitem[Berns et~al.(2025)Berns, Rodrigues, Finocchio, and Mentink]{berns2025predicting}
Rutger J. L.~F. Berns, Davi~R. Rodrigues, Giovanni Finocchio, and Johan~H. Mentink.
\newblock Predicting sampling advantage of stochastic ising machines for quantum simulations, 2025.

\bibitem[Brahma et~al.(2025)Brahma, Han, Razzaque, Patel, and Salahuddin]{brahma2025hardware}
Pratik Brahma, Junghoon Han, Tamzid Razzaque, Saavan Patel, and Sayeef Salahuddin.
\newblock Hardware acceleration of frustrated lattice systems using convolutional restricted boltzmann machine, 2025.

\bibitem[Aadit et~al.(2022)Aadit, Grimaldi, Carpentieri, Theogarajan, Martinis, Finocchio, and Camsari]{aadit2022massively}
Navid~Anjum Aadit, Andrea Grimaldi, Mario Carpentieri, Luke Theogarajan, John~M Martinis, Giovanni Finocchio, and Kerem~Y Camsari.
\newblock {Massively parallel probabilistic computing with sparse Ising machines}.
\newblock \emph{Nature Electronics}, 5\penalty0 (7):\penalty0 460--468, 2022.

\bibitem[Niazi et~al.(2024)Niazi, Chowdhury, Aadit, Mohseni, Qin, and Camsari]{niazi_training_2024}
Shaila Niazi, Shuvro Chowdhury, Navid~Anjum Aadit, Masoud Mohseni, Yao Qin, and Kerem~Y. Camsari.
\newblock Training deep {Boltzmann} networks with sparse {Ising} machines.
\newblock \emph{Nature Electronics}, pages 1--10, June 2024.
\newblock Publisher: Nature Publishing Group.

\bibitem[Fedorovich et~al.(2025)Fedorovich, Devos, Haegeman, Vanderstraeten, Verstraete, and Ueda]{fedorovich2025finite}
Gleb Fedorovich, Lukas Devos, Jutho Haegeman, Laurens Vanderstraeten, Frank Verstraete, and Atsushi Ueda.
\newblock Finite-size scaling on the torus with periodic projected entangled-pair states.
\newblock \emph{Phys. Rev. B}, 111:\penalty0 165124, Apr 2025.

\bibitem[King et~al.(2021)King, Raymond, Lanting, Isakov, Mohseni, Poulin-Lamarre, Ejtemaee, Bernoudy, Ozfidan, Smirnov, Reis, Altomare, Babcock, Baron, Berkley, Boothby, Bunyk, Christiani, Enderud, Evert, Harris, Hoskinson, Huang, Jooya, Khodabandelou, Ladizinsky, Li, Lott, MacDonald, Marsden, Marsden, Medina, Molavi, Neufeld, Norouzpour, Oh, Pavlov, Perminov, Prescott, Rich, Sato, Sheldan, Sterling, Swenson, Tsai, Volkmann, Whittaker, Wilkinson, Yao, Neven, Hilton, Ladizinsky, Johnson, and Amin]{king2021scaling}
Andrew~D. King, Jack Raymond, Trevor Lanting, Sergei~V. Isakov, Masoud Mohseni, Gabriel Poulin-Lamarre, Sara Ejtemaee, William Bernoudy, Isil Ozfidan, Anatoly~Yu. Smirnov, Mauricio Reis, Fabio Altomare, Michael Babcock, Catia Baron, Andrew~J. Berkley, Kelly Boothby, Paul~I. Bunyk, Holly Christiani, Colin Enderud, Bram Evert, Richard Harris, Emile Hoskinson, Shuiyuan Huang, Kais Jooya, Ali Khodabandelou, Nicolas Ladizinsky, Ryan Li, P.~Aaron Lott, Allison J.~R. MacDonald, Danica Marsden, Gaelen Marsden, Teresa Medina, Reza Molavi, Richard Neufeld, Mana Norouzpour, Travis Oh, Igor Pavlov, Ilya Perminov, Thomas Prescott, Chris Rich, Yuki Sato, Benjamin Sheldan, George Sterling, Loren~J. Swenson, Nicholas Tsai, Mark~H. Volkmann, Jed~D. Whittaker, Warren Wilkinson, Jason Yao, Hartmut Neven, Jeremy~P. Hilton, Eric Ladizinsky, Mark~W. Johnson, and Mohammad~H. Amin.
\newblock Scaling advantage over path-integral monte carlo in quantum simulation of geometrically frustrated magnets.
\newblock \emph{Nature Communications}, 12\penalty0 (1):\penalty0 1113, 02 2021.

\bibitem[Sachdev(2011)]{sachdev2011quantum}
Subir Sachdev.
\newblock \emph{Quantum Phase Transitions}.
\newblock Cambridge University Press, 2 edition, 2011.

\bibitem[Marshall(1955)]{marshall1955anti}
W.~Marshall.
\newblock Antiferromagnetism.
\newblock \emph{Proceedings of the Royal Society of London. A. Mathematical and Physical Sciences}, 232\penalty0 (1188):\penalty0 48--68, 10 1955.

\bibitem[Bravyi et~al.(2008)Bravyi, Divincenzo, Oliveira, and Terhal]{bravyi2008complexity}
Sergey Bravyi, David~P. Divincenzo, Roberto Oliveira, and Barbara~M. Terhal.
\newblock The complexity of stoquastic local hamiltonian problems.
\newblock \emph{Quantum Info. Comput.}, 8\penalty0 (5):\penalty0 361–385, May 2008.

\bibitem[Szab\'o and Castelnovo(2020)]{szabo2020neural}
Attila Szab\'o and Claudio Castelnovo.
\newblock Neural network wave functions and the sign problem.
\newblock \emph{Phys. Rev. Res.}, 2:\penalty0 033075, Jul 2020.

\bibitem[Torlai et~al.(2018)Torlai, Mazzola, Carrasquilla, Troyer, Melko, and Carleo]{Torlai2018neural}
Giacomo Torlai, Guglielmo Mazzola, Juan Carrasquilla, Matthias Troyer, Roger Melko, and Giuseppe Carleo.
\newblock Neural-network quantum state tomography.
\newblock \emph{Nature Physics}, 14\penalty0 (5):\penalty0 447--450, May 2018.

\bibitem[Chowdhury et~al.(2025)Chowdhury, Aadit, Grimaldi, Raimondo, Raut, Lott, Mentink, Rams, Ricci-Tersenghi, Chiappini, Theogarajan, Srimani, Finocchio, Mohseni, and Camsari]{chowdhury2025pushing}
Shuvro Chowdhury, Navid~Anjum Aadit, Andrea Grimaldi, Eleonora Raimondo, Atharva Raut, P.~Aaron Lott, Johan~H. Mentink, Marek~M. Rams, Federico Ricci-Tersenghi, Massimo Chiappini, Luke~S. Theogarajan, Tathagata Srimani, Giovanni Finocchio, Masoud Mohseni, and Kerem~Y. Camsari.
\newblock Pushing the boundary of quantum advantage in hard combinatorial optimization with probabilistic computers.
\newblock \emph{Nature Communications}, 16\penalty0 (1):\penalty0 9193, Oct 2025.

\bibitem[Deng et~al.(2017)Deng, Li, and Das~Sarma]{deng2017machine}
Dong-Ling Deng, Xiaopeng Li, and S.~Das~Sarma.
\newblock Machine learning topological states.
\newblock \emph{Phys. Rev. B}, 96:\penalty0 195145, Nov 2017.

\bibitem[Nomura et~al.(2017)Nomura, Darmawan, Yamaji, and Imada]{nomura2017restricted}
Yusuke Nomura, Andrew~S. Darmawan, Youhei Yamaji, and Masatoshi Imada.
\newblock Restricted boltzmann machine learning for solving strongly correlated quantum systems.
\newblock \emph{Phys. Rev. B}, 96:\penalty0 205152, Nov 2017.

\bibitem[Blackman and Vigna(2021)]{blackman2021scrambled}
David Blackman and Sebastiano Vigna.
\newblock Scrambled linear pseudorandom number generators.
\newblock \emph{ACM Trans. Math. Softw.}, 47\penalty0 (4), September 2021.

\bibitem[Chowdhury et~al.(2023)Chowdhury, Camsari, and Datta]{chowdhury_accelerated_2023}
Shuvro Chowdhury, Kerem~Y. Camsari, and Supriyo Datta.
\newblock Accelerated quantum {Monte} {Carlo} with probabilistic computers.
\newblock \emph{Communications Physics}, 6\penalty0 (1):\penalty0 1--9, April 2023.
\newblock Number: 1 Publisher: Nature Publishing Group.

\bibitem[Sorella(1998)]{sorella1998green}
Sandro Sorella.
\newblock Green function monte carlo with stochastic reconfiguration.
\newblock \emph{Phys. Rev. Lett.}, 80:\penalty0 4558--4561, May 1998.

\bibitem[Neuscamman et~al.(2012)Neuscamman, Umrigar, and Chan]{neuscamman2012optimizing}
Eric Neuscamman, C.~J. Umrigar, and Garnet Kin-Lic Chan.
\newblock Optimizing large parameter sets in variational quantum monte carlo.
\newblock \emph{Phys. Rev. B}, 85:\penalty0 045103, Jan 2012.

\bibitem[Nikhar et~al.(2024)Nikhar, Kannan, Aadit, Chowdhury, and Camsari]{nikhar2024all}
Srijan Nikhar, Sidharth Kannan, Navid~Anjum Aadit, Shuvro Chowdhury, and Kerem~Y Camsari.
\newblock {All-to-all reconfigurability with sparse and higher-order Ising machines}.
\newblock \emph{Nature Communications}, 15\penalty0 (1):\penalty0 8977, 2024.

\bibitem[Bl\"ote and Deng(2002)]{blote2002cluster}
Henk W.~J. Bl\"ote and Youjin Deng.
\newblock Cluster monte carlo simulation of the transverse ising model.
\newblock \emph{Phys. Rev. E}, 66:\penalty0 066110, Dec 2002.

\bibitem[Gao and Duan(2017)]{gao2017efficient}
Xun Gao and Lu-Ming Duan.
\newblock Efficient representation of quantum many-body states with deep neural networks.
\newblock \emph{Nature Communications}, 8\penalty0 (1):\penalty0 662, Sep 2017.

\bibitem[Carleo et~al.(2018)Carleo, Nomura, and Imada]{carleo2018constructing}
Giuseppe Carleo, Yusuke Nomura, and Masatoshi Imada.
\newblock Constructing exact representations of quantum many-body systems with deep neural networks.
\newblock \emph{Nature Communications}, 9\penalty0 (1):\penalty0 5322, Dec 2018.

\bibitem[Salakhutdinov and Hinton(2009)]{salakhutdinov2009deep}
Ruslan Salakhutdinov and Geoffrey~E. Hinton.
\newblock Deep boltzmann machines.
\newblock In David A.~Van Dyk and Max Welling, editors, \emph{Proceedings of the Twelfth International Conference on Artificial Intelligence and Statistics, {AISTATS} 2009, Clearwater Beach, Florida, USA, April 16-18, 2009}, volume~5 of \emph{{JMLR} Proceedings}, pages 448--455. JMLR.org, 2009.

\bibitem[Nomura et~al.(2021)Nomura, Yoshioka, and Nori]{nomura_purifying_2021}
Yusuke Nomura, Nobuyuki Yoshioka, and Franco Nori.
\newblock Purifying {Deep} {Boltzmann} {Machines} for {Thermal} {Quantum} {States}.
\newblock \emph{Physical Review Letters}, 127\penalty0 (6):\penalty0 060601, August 2021.
\newblock Publisher: American Physical Society.

\bibitem[Hinton(2002)]{hinton2002training}
Geoffrey~E. Hinton.
\newblock Training products of experts by minimizing contrastive divergence.
\newblock \emph{Neural Computation}, 14\penalty0 (8):\penalty0 1771--1800, August 2002.

\bibitem[Hinton et~al.(2006)Hinton, Osindero, and Teh]{hinton2006fast}
Geoffrey~E. Hinton, Simon Osindero, and Yee-Whye Teh.
\newblock A fast learning algorithm for deep belief nets.
\newblock \emph{Neural Computation}, 18\penalty0 (7):\penalty0 1527--1554, July 2006.

\bibitem[Bell and Garland(2009)]{bell2009implementing}
Nathan Bell and Michael Garland.
\newblock Implementing sparse matrix--vector multiplication on throughput-oriented processors.
\newblock \emph{Proceedings of the Conference on High Performance Computing Networking, Storage and Analysis (SC)}, 2009.

\bibitem[Williams et~al.(2009)Williams, Waterman, and Patterson]{williams2009optimization}
Samuel Williams, Andrew Waterman, and David Patterson.
\newblock Optimization of sparse matrix--vector multiplication on emerging multicore platforms.
\newblock \emph{Parallel Computing}, 35\penalty0 (3):\penalty0 178--194, 2009.

\bibitem[Chen et~al.(2018)Chen, Satish, Hong, Oguntebi, and Olukotun]{chen2018gunrock}
Yangzihao Chen, Nadathur Satish, Sungpack Hong, Oluwasegun Oguntebi, and Kunle Olukotun.
\newblock Gunrock: Gpu graph analytics.
\newblock \emph{ACM Transactions on Parallel Computing}, 4\penalty0 (1):\penalty0 1--49, 2018.

\bibitem[Horowitz(2014)]{horowitz2014computing}
Mark Horowitz.
\newblock 1.1 computing's energy problem (and what we can do about it).
\newblock \emph{IEEE International Solid-State Circuits Conference (ISSCC)}, pages 10--14, 2014.

\bibitem[Gale et~al.(2019)Gale, Elsen, and Hooker]{gale2019state}
Trevor Gale, Erich Elsen, and Sara Hooker.
\newblock The state of sparsity in deep neural networks.
\newblock \emph{arXiv preprint arXiv:1902.09574}, 2019.

\bibitem[Karypis and Kumar(1998)]{karypis1998fast}
George Karypis and Vipin Kumar.
\newblock A fast and high quality multilevel scheme for partitioning irregular graphs.
\newblock \emph{SIAM Journal on Scientific Computing}, 20\penalty0 (1):\penalty0 359--392, 1998.

\end{thebibliography}

\setcounter{secnumdepth}{3}

\newcommand{\beginsupplement}{%
        \ifdefined\counterwithout
            \counterwithout{equation}{section}
        \fi
        \setcounter{table}{0}
        \renewcommand{\thetable}{S\arabic{table}}%
        \setcounter{figure}{0}
        \renewcommand{\thefigure}{S\arabic{figure}}%
        \setcounter{equation}{0}
        \renewcommand{\theequation}{S.\arabic{equation}}%
        \renewcommand{\thealgocf}{S\arabic{algocf}}
        \setcounter{algocf}{0}%
     }

\onecolumngrid
\begin{center}
{\sffamily\Large\bf Supplementary Information\par}
\vskip 0.5em
{\sffamily\LARGE\bf Probabilistic Computers for Neural Quantum States \par}
\vspace{1em}
\normalfont\noindent{\sffamily Shuvro Chowdhury, Jasper Pieterse, Navid Anjum Aadit, Shaila Niazi, Johan H. Mentink, and Kerem Y. Camsari}
\end{center}

\beginsupplement
\renewcommand{\theHfigure}{S\arabic{figure}}
\renewcommand{\theHtable}{S\arabic{table}}
\renewcommand{\theHequation}{S.\arabic{equation}}
\renewcommand{\theHalgocf}{S\arabic{algocf}}
\renewcommand{\theHsection}{S\arabic{section}}

\setcounter{section}{0}
\setcounter{subsection}{0}

\vspace{-15pt}

\section{Runtime Profiling and Bottleneck Analysis for 2D TFIM}
\label{supp_sec:runtime_profiling}

To quantify the removal of the sampling bottleneck, we profiled the wall-clock execution time of the training loop across three hardware configurations for 2D TFIM with $L = 35$: a standard CPU implementation (11th Gen Intel Core i7-11700 @ 2.50 GHz, 96 GB RAM: also used in the hybrid setup), a MATLAB-based high-performance GPU implementation (NVIDIA GeForce RTX 4060 Ti, 16 GB VRAM), and our hybrid CPU-FPGA setup.

\begin{figure}[!ht]
    \centering
    \includegraphics[width=3.5in,keepaspectratio]{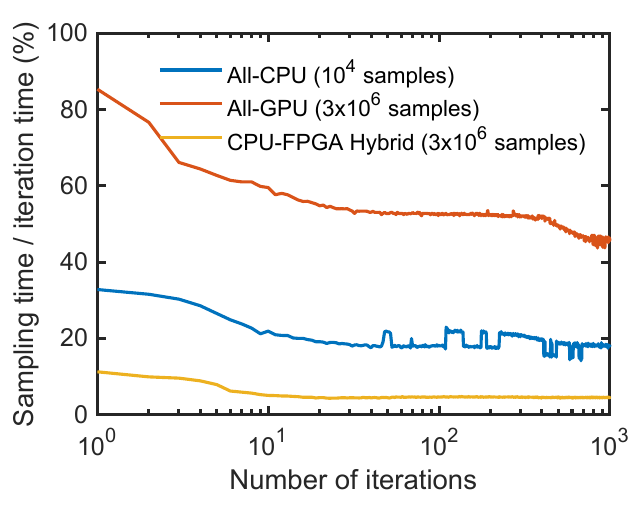} 
    \caption{\textbf{Evolution of the computational bottleneck.} For 2D TFIM of size $L = 35$, the fraction of total iteration time spent on Monte Carlo sampling is plotted against training iterations for three architectures. The \textbf{All-GPU} setup (red, $3\times10^6$ samples) remains dominated by sampling costs ($>50\%$), indicating a throughput bottleneck. The \textbf{All-CPU} baseline (blue) spends $\sim20-30\%$ of its time on sampling despite processing $300\times$ fewer samples ($10^4$, for a manageable end-to-end training).  In contrast, the \textbf{CPU-FPGA Hybrid} (yellow) performs the massive $3\times10^6$ sampling task in less than $5\%$ of the total loop time, effectively rendering the sampling step instantaneous relative to the classical gradient accumulation.}
    \label{fig:bottleneck_comparison}
\end{figure}

\section{Implementation Details of the Multi-FPGA System}
\label{supp_sec:details_cluster}
\begin{figure*}[!ht]
    \centering \includegraphics[width=0.95\textwidth,keepaspectratio]{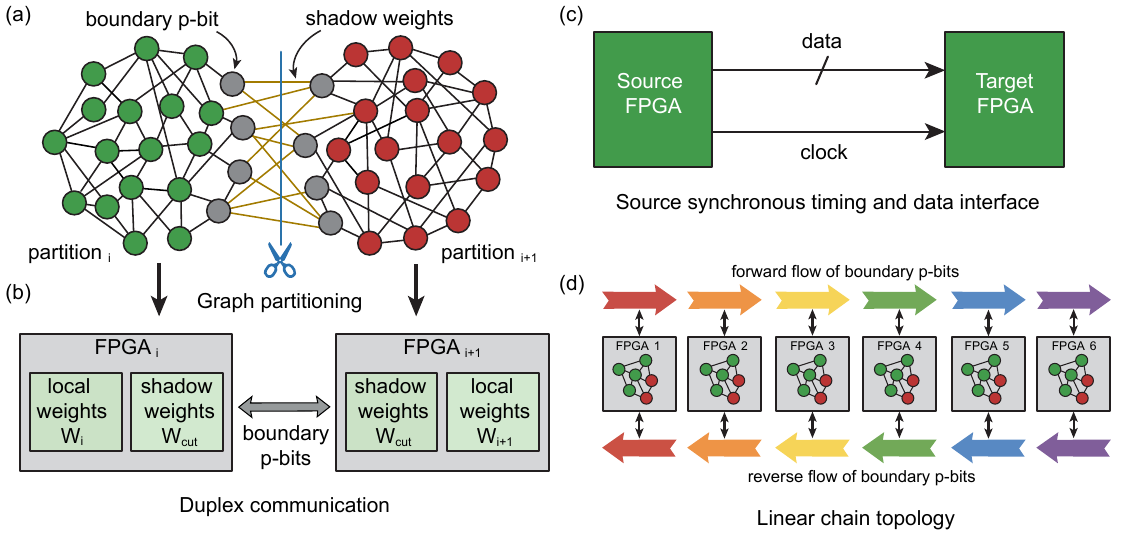} 
    \caption{\textbf{Multi-FPGA cluster architecture for large-scale NQS.} (a) A sparse Ising graph is partitioned into subgraphs using the METIS min-cut graph partitioner~\cite{karypis1998fast} and the subgraphs are mapped to neighboring FPGAs. (b) Each FPGA stores its local weights and duplicates cross-partition couplings as \emph{shadow weights}. Only binary boundary p-bit states are exchanged over full-duplex FMC links (both directions) since the coupling matrix is symmetric. (c) Source-synchronous boundary interface: clock and data are forwarded together. Incoming boundary p-bits remain asynchronous to the destination FPGA clock domain and the forwarded clock is used only to align the received p-bit stream. (d) A 6-FPGA linear chain illustrating forward and reverse flow of boundary p-bit states over nearest-neighbor links.}
    \label{supp_fig:cluster_arch}
\end{figure*}

We provide an extended view of the cluster architecture in Fig.~\ref{supp_fig:cluster_arch}, specifically detailing the mapping of partitioned subgraphs to hardware and the  protocols used for inter-FPGA data exchange.

\section{Detailed Dual Sampling Algorithm}
Here, we provide a more elaborative version of the dual sampling algorithm.

\SetKwComment{Comment}{\normalfont $\triangleright$ }{}
\SetKwInput{KwInput}{Input}                
\SetKwInput{KwOutput}{Output}              
\SetKwFor{ParFor}{for}{do in parallel}{end}

\begin{algorithm}[!ht]
    \DontPrintSemicolon
    \KwInput{Number of qubits ($N_Q$), parameters of the quantum Hamiltonian ($J, h, \Gamma$), number of samples per iterations ($N_s$), number of clamped samples ($N_c$), number of iterations ($N_{\mathrm{iter}}$), learning rate schedule ($\eta$), optimization parameters}
    
    \KwOutput{Optimized parameters: biases ($b$), couplings ($W$), ground energy estimate ($E_\mathrm{G}$)}
        Define DBM adjacency matrix, $A$.\;
        Initialize biases: $b_i \gets 0.01\, \mathcal{N}(0,1)$, \quad  $\forall\,  i \in \{v,h,d\}$.\;
        Initialize weights: $W_{ij} \gets 0.01\, \mathcal{N}(0,1)$\;
        Initialize p-bit states randomly from $\{-1,+1\}$.\; 
        
        \For{each iteration $t = 1$ to $N_{\mathrm{iter}}$}{
            \For{each sample in $N_s$}{
            \ParFor{each spin in visible layer}{
                Compute input $I_i=\sum_{j}W_{ij}\sigma_j+h_i$.\;
                Update:  $\sigma_i=\text{sgn}\left(\tanh(\beta I_i)-r_{[-1,1]}\right)$.\;
            }
            Clamp or fix the spins in the visible layer, $v$.\;
            Compute diagonal contribution $H_{v,v}$.\;
            $E_{\rm loc}(v) \gets H_{v,v}$.\;
            $O_{v_i} \gets O_{v_i} + 0.5\,v_i$, \quad $\forall\,  i \in v$.\;
            $O_{h_j}(v) \gets 0$, $O_{d_k}(v) \gets 0$, $O_{W_{ij}}(v) \gets 0$.\; 
            \For{each sample in $N_c$}{
                \ParFor{each spin in hidden layer}{
                    Compute input $I_i=\sum_{j}W_{ij}\sigma_j+h_i$.\;
                    Update:  $\sigma_i=\text{sgn}\left(\tanh(\beta I_i)-r_{[-1,1]}\right)$.\;
                }
                $O_{h_j}(v) \gets O_{h_j}(v) + 0.5\,h_j$, \quad $\forall\,  j \in h$.\;
                \ParFor{each spin in deep layer}{
                    Compute input $I_i=\sum_{j}W_{ij}\sigma_j+h_i$.\;
                    Update:  $\sigma_i=\text{sgn}\left(\tanh(\beta I_i)-r_{[-1,1]}\right)$.\;
                }
                $O_{d_k}(v) \gets O_{d_k}(v) + 0.5\,d_k$, \quad $\forall\,  k \in d$.\;

                Compute input $I_i$, \quad  $\forall\,  i \in v$.\;
                $p_{\text{flip},i} \gets p_{\text{flip},i} + \exp{(-2I_iv_i)}$, \quad  $\forall\,  i \in v$.\;
                $p_{\text{flip\_sq},i} \gets p_{\text{flip\_sq},i} + \exp{(-4I_iv_i)}$, \quad  $\forall\,  i \in v$.\;
                $O_{W_{ij}}(v) \gets O_{W_{ij}}(v) + 0.5\,\sigma_i\sigma_j$, \, $\forall\,  i,j \in \{v,h,d\}$.\;
                }
                $O_{h_j}(v) \gets O_{h_j}(v)/N_c$, \quad $\forall\,  j \in h$.\;
                $O_{d_k}(v) \gets O_{d_k}(v)/N_c$, \quad $\forall\,  j \in d$.\;
                $O_{W_{ij}}(v) \gets O_{W_{ij}}(v)/N_c$, \, $\forall\,  i,j \in \{v,h,d\}$.\;
                $p_{\text{flip},i} \gets p_{\text{flip},i}/N_c$, \quad  $\forall\,  i \in v$.\;
                $p_{\text{flip\_sq},i} \gets p_{\text{flip\_sq},i} / N_c$, \quad  $\forall\,  i \in v$.\;
                $\text{Var}_{\text{pop},i} \gets (p_{\text{flip\_sq},i} - (p_{\text{flip},i})^2)/N_c$ \quad  $\forall\,  i \in v$.\;
                $\Delta_i \gets \text{Var}_{\text{pop},i}/(8 p_{\text{flip},i} \sqrt{p_{\text{flip},i}})$ \quad  $\forall\,  i \in v$.\;
                \For{each spin in visible layer}{
                    Get $v^{(i)}$ by flipping the spin.\;
                    $E_{\rm loc}(v) \gets E_{\rm loc}(v) + H_{v,v^{(i)}}(\sqrt{p_{\text{flip},i}}+\Delta_i)$.\;
                }
            }
            Use SR for gradients $\Delta b_i$ and $\Delta W_{ij}$, \, $\forall\,  i,j \in \{v,h,d\}$.\;  
            Update biases: $b_i \gets b_i - \eta(t)\Delta\,b_i$, \, $\forall\,  i \in \{v,h,d\}$.\;
            Update weights: $W_{ij} \gets W_{ij} - \eta(t)\Delta W_{ij}$, \, $\forall\,  i,j \in \{v,h,d\}$.\;
        }
        \KwRet Optimized $b$, $W$ and $E_{\mathrm{G}}^{(N_{\mathrm{iter}})}$.\;
    
    \caption{Algorithm for Machine Learning Quantum Hamiltonians using deep NQS}
    \label{algo:fullDualSampling}
\end{algorithm}

\section{Exactness of the dual-sampling estimator in the infinite-sample limit}
\label{supp_sec:proof_dualSampling}

\noindent Let us define a neural quantum state  from a probabilistic DBM model \(p_{\theta}(v)\) over spin configurations \(v\in\{\pm 1\}^N\) by
\begin{align}
\Psi_{\theta}(v) = \sqrt{p_{\theta}(v)},\qquad \text{where} \quad p_{\theta}(v)=\sum_{h,d}p_{\theta}(v,h,d), \quad \text{and} \quad p_{\theta}(v,h,d)=\frac{1}{Z_{\theta}}\,e^{-E_{\theta}(v,h,d)}
\label{eq:proof1}
\end{align}
so that \(|\Psi_{\theta}(v)|^2 = p_{\theta}(v)\). Consider the transverse-field Ising Hamiltonian
\begin{align}
H = -\sum_{\langle ij\rangle} J_{ij}\hat{\sigma}_i^z\hat{\sigma}_j^z
    - \Gamma_x \sum_{i=1}^{N} \hat{\sigma}_i^x \nonumber
\end{align}
where the sum over $\langle ij \rangle$ includes only unique pairs connected by an edge in the underlying lattice. In the following, we will assume that
\begin{enumerate}
\item[(1)] in outer sampling,  we can draw independent and identically distributed (i.i.d.) visible configurations from \(p_{\theta}(v)\);
\item[(2)] in inner conditional sampling, for any fixed visible configuration \(v\), we can sample the hidden/deep variables from their exact conditionals using the DBM and obtain an unbiased estimator of the ratio \(p_{\theta}(v^{(i)})/p_{\theta}(v)\) for every single-spin flip \(v^{(i)}\) (the configuration obtained from \(v\) by flipping spin \(i\));
\end{enumerate}

\noindent  For any state \(\Psi_{\theta}\), the exact variational energy is defined as
\begin{align}
E(\theta) \;=\;
\frac{\langle \Psi_{\theta} | H | \Psi_{\theta} \rangle}
     {\langle \Psi_{\theta} | \Psi_{\theta} \rangle} \nonumber
\end{align}
\noindent  The usual
variational Monte Carlo identity is
\begin{equation}
E(\theta)
= \frac{\sum_{v}{|\Psi_{\theta}(v)|^2 \, E_{\text{loc}}(v)}}{\sum_{v}{|\Psi_{\theta}(v)|^2}}
= \mathbb{E}_{v\sim |\Psi_{\theta}|^2}\bigl[ E_{\text{loc}}(v) \bigr],
\label{eq:proof2}
\end{equation}
where the \emph{local energy} is
\begin{equation}
E_{\text{loc}}(v) = \frac{(H\Psi_{\theta})(v)}{\Psi_{\theta}(v)}
\label{eq:proof3}
\end{equation}

\noindent For the TFIM Hamiltonian above, in the \(\sigma^z\) basis, we have
\begin{equation}
E_{\text{loc}}(v)
= -\sum_{\langle ij\rangle} J_{ij}\sigma_i^z\sigma_j^z
  - \Gamma_x \sum_{i=1}^{N}
      \frac{\Psi_{\theta}(v^{(i)})}{\Psi_{\theta}(v)} 
\label{eq:proof4}
\end{equation}

\noindent Now, by construction of the NQS,
\begin{equation}
\Psi_{\theta}(v) = \sqrt{p_{\theta}(v)}
\quad\Longrightarrow\quad
\frac{\Psi_{\theta}(v^{(i)})}{\Psi_{\theta}(v)}
= \sqrt{\frac{p_{\theta}(v^{(i)})}{p_{\theta}(v)}}=\sqrt{r_i(v)}
\label{eq:proof4_2}
\end{equation}
So if, for a fixed visible configuration \(v\), we can obtain an unbiased estimate of \(\frac{p_{\theta}(v^{(i)})}{p_{\theta}(v)}\), which is exactly what the inner (hidden/deep) sampling does in Algorithm~\ref{algo:fullDualSampling},  then applying the square root exactly recovers the amplitude ratio in Eq.~(\ref{eq:proof4_2}). Let us denote the inner estimate (after averaging over \(N_c\) conditional samples) by \(\hat r_i(v)\). Under assumption~(2) and letting \(N_c \to \infty\), we can show that
\begin{equation}
\lim_{N_c\to\infty} \hat r_i(v) = \frac{p_{\theta}(v^{(i)})}{p_{\theta}(v)}
\quad\Longrightarrow\quad
\lim_{N_c\to\infty} \sqrt{\hat r_i(v)} =
\sqrt{\frac{p_{\theta}(v^{(i)})}{p_{\theta}(v)}}
= \frac{\Psi_{\theta}(v^{(i)})}{\Psi_{\theta}(v)} .
\label{eq:proof5}
\end{equation}

\noindent Plugging this into Eq.~(\ref{eq:proof4}), we obtain an \emph{inner-level} local-energy
estimator
\begin{equation}
\lim_{N_c\to\infty}\hat E_{\text{loc}}(v)
= -\sum_{\langle ij\rangle} J_{ij}\sigma_i^z\sigma_j^z - \Gamma_x \sum_{i=1}^{N} \left[\lim_{N_C\to\infty} \sqrt{\hat r_i(v)}\right]
= E_{\text{loc}}(v)
\label{eq:proof6}
\end{equation}

To see why Eq.~(\ref{eq:proof6}) holds, consider a visible configuration $v$ and the configuration $v^{(i)}$ obtained by flipping spin $v_i$. For fixed hidden and deep spins $(h,d)$, the DBM defines a Boltzmann weight 
\begin{align}
p_{\theta}(v,h,d) = \frac{1}{Z_{\theta}}\, e^{-E_{\theta}(v,h,d)}
\label{eq:proof7}
\end{align}
and
\begin{align}
\Delta\,E_{i}(v,h,d)=E_{\theta}(v^{(i)},h,d)-E_{\theta}(v,h,d)=2I_i(v,h,d)v_i
\label{eq:proof8}
\end{align}

\noindent so the ratio of joint probabilities is
\begin{align}
\frac{p_{\theta}(v^{(i)},h,d)}{p_{\theta}(v,h,d)} = e^{-\Delta\,E_{i}(v,h,d)} = e^{-2I_i(v,h,d)v_i}
\label{eq:proof9}
\end{align}
The marginal probability of a visible configuration is obtained by summing over hidden and deep variables:
\begin{align}
p_{\theta}(v) = \sum_{h,d}{p_{\theta}(v,h,d)}, \qquad p_{\theta}(v^{(i)}) = \sum_{h,d}{p_{\theta}(v^{(i)},h,d)}
\label{eq:proof10}
\end{align}

\noindent Using Eq.~(\ref{eq:proof9}) in the second expression gives
\begin{align}
p_{\theta}(v^{(i)}) = \sum_{h,d}{p_{\theta}(v^{(i)},h,d)} = \sum_{h,d}{p_{\theta}(v,h,d)}\,e^{-\Delta\,E_{i}(v,h,d)}
\label{eq:proof11}
\end{align}
Therefore,
\begin{align}
\frac{p_{\theta}(v^{(i)})}{p_{\theta}(v)} = \frac{\sum_{h,d}{p_{\theta}}(v^{(i)},h,d)}{\sum_{h,d}{p_{\theta}(v,h,d)}} = \frac{\sum_{h,d}{p_{\theta}}(v,h,d)\,e^{-\Delta\,E_{i}(v,h,d)}}{\sum_{h,d}{p_{\theta}(v,h,d)}} = \mathbb{E}_{(h,d)\sim p(h,d|v)}\bigl[ e^{-\Delta\,E_{i}(v,h,d)} \bigr]=r_i(v)
\label{eq:proof12}
\end{align}
which establishes Eq.~(\ref{eq:proof6}). $r_i(v)$ is the conditional expectation, under the clamped distribution $p_{\theta}(h,d|v)$, of the joint Boltzmann ratio between $v^{(i)}$ and $v$. In our algorithm, this expectation is estimated by a Monte Carlo average over $N_c$ clamped samples,
\begin{align}
\hat{r}_i(v) = \frac{1}{N_c}\sum_{k=1}^{N_c}{e^{-\Delta\,E_{i}(v,h^{(k)},d^{(k)})}}
\end{align}
and by the law of large numbers $\hat{r}_i(v)\to r_i(v)$ as $N_c\to\infty$, so the estimator converges to the exact marginal ratio $\displaystyle\frac{p_{\theta}(v^{(i)})}{p_{\theta}(v)}$.

\section{Training Hyperparameters and Optimization Details}
\label{supp_sec:hyperparamters}

All Neural Quantum State (NQS) models presented in this work were trained using a custom MATLAB framework accelerated by CUDA-enabled GPUs. The optimization was performed using Stochastic Reconfiguration (SR) implemented via an iterative Conjugate Gradient (CG) solver to avoid explicitly constructing the dense Fisher Information Matrix (FIM).

\subsection{Hyperparameters}
The specific hyperparameters used for the training of the Sparse Deep Boltzmann Machine (DBM) and Sparse RBM results shown in Figs.~\ref{fig:fig4}~and~\ref{fig:fig5} are summarized in Table~\ref{tab:hyperparams}.

\begin{itemize}
    \item \textbf{Dual Sampling Scheme:} For the Sparse DBM, the gradient estimation involves two distinct sampling populations. The \emph{outer loop} uses $N_s = 10,000$ samples to estimate expectations over the visible layer $|\Psi(v)|^2$. The \emph{inner loop} (conditional sampling) uses $N_c = 1,000$ clamped samples to estimate the gradients of the deep layers and the local energy ratios. For a fixed visible configuration, the same set of $N_c$ conditional samples is reused to estimate all single-spin-flip ratios, such that each ratio estimator is based on $N_c$ samples (not $N_c/N$).
    \item \textbf{Learning Rate Schedule:} For the algorithmic benchmarks in Fig.~4, we employed a time-dependent learning rate $\eta(t)$ following a \emph{cosine decay schedule}, decreasing from $\eta_{\rm max} = 0.1$ to $\eta_{\rm  min} = 10^{-5}$ over the course of training to ensure stable convergence. Conversely, for the hardware demonstrations in Fig.~2 and Fig.~3, a cosine decay schedule from $\eta_{\rm max} = 0.05$ to $\eta_{\rm min} = 0.01$ was employed.
    \item \textbf{Regularization:} To stabilize the inversion of the Fisher Information Matrix, we applied a diagonal shift regularization (Tikhonov regularization) $S \rightarrow S + \lambda I$. The shift parameter $\lambda$ was initialized at $\lambda_0 = 0.1$ and decayed adaptively during training to a minimum of $\lambda_{\rm min} = 10^{-4}$ using a decay factor $b_0 = 0.9$ per iteration ($\lambda_t = \text{max}(\lambda_{\rm min}, \lambda_0 b_0^t$).
\end{itemize}

The outer sampling budget $N_s$ and the inner conditional sampling budget $N_c$ control distinct sources of statistical error. The outer loop samples the full visible-spin distribution and determines the conditioning of the stochastic-reconfiguration update, while the inner loop estimates conditional expectations over auxiliary variables for a fixed visible configuration. Because the conditional distributions are lower-variance and strongly constrained, accurate estimation of wavefunction ratios can be achieved with substantially fewer conditional samples. In practice, we find that $N_c = 10^3$ is sufficient to stabilize training and achieve chemical accuracy for the systems studied here. We do not claim that this choice is asymptotically optimal. In general, the hyperparameters listed in Table S1 denote fixed sampling budgets used throughout this work. We do not assume or claim that these quantities are independent of the system size. Rather, they are chosen empirically to ensure stable optimization and accurate energy estimation for the system sizes studied here. Determining how these sampling budgets must scale with system size to maintain a prescribed statistical accuracy would require a dedicated variance analysis and is beyond the scope of the present work.

\subsection{Optimization Routine}
The model parameters were updated using the Stochastic Reconfiguration (SR) method. Instead of inverting the curvature matrix $S$ directly (which scales as $\mathcal{O}(N_p^3)$), we solved the linear system $S \cdot \delta \theta = g$ using a Preconditioned Conjugate Gradient (PCG) solver.

\begin{itemize}
    \item \textbf{Matrix-Free Implementation:} The solver utilizes implicit matrix-vector products to compute $S \cdot v$ without ever instantiating the full $S$ matrix in memory.
    \item \textbf{CG Tolerances:} The linear solver was configured with a relative tolerance of $10^{-4}$ and a maximum of 500 iterations per optimization step.
\end{itemize}

\subsection{Final Evaluation}
Following the training phase (1,000 iterations), the optimized model parameters were frozen. A final evaluation run was performed using a significantly larger sample size of $N_{\rm eval} = 10^6$ to obtain the high-precision energy estimates and error bars reported in the main text figures.

\begin{table}[!ht]
\centering
\caption{Summary of Training Hyperparameters}
\label{tab:hyperparams}
\begin{tabular}{@{}lccc@{}}
\hline \hline
\textbf{Parameter} & \textbf{Symbol} & \textbf{Value} & \textbf{Description} \\
\hline
Training Duration & $N_{\rm iter}$ & 1000 & Total optimization steps \\
Visible Samples & $N_s$ & $10,000$ & Samples drawn from $|\Psi(v)|^2$ (Outer loop) \\
Clamped Samples & $N_c$ & $1,000$ & Conditional samples for Dual Sampling (Inner loop) \\
Initial Learning Rate & $\eta_{\rm max}$ & $0.1$ & Maximum learning rate \\
Final Learning Rate & $\eta_{\rm min}$ & $10^{-5}$ & Minimum learning rate \\
SR Diagonal Shift & $\lambda_0$ & $0.1$ & Initial regularization for Fisher Matrix \\
SR $\lambda$ decay factor & $b_0$ & $0.9$ & Geometric decay rate for diagonal shift \\
CG Tolerance & $\epsilon_{\rm CG}$ & $10^{-4}$ & Relative tolerance for linear solver \\
CG Max Iterations & $k_{\rm CG}$ & 500 & Max iterations for linear solver \\
Evaluation Samples & $N_{\rm eval}$ & $10^6$ & Samples used for final energy estimation \\
\hline \hline
\end{tabular}
\end{table}

\section{Parameter Counting and Sparse Connectivity Definitions}
\label{supp_sec:parameter_counting}

To quantify the representational efficiency of Deep Boltzmann Machines (DBMs) against Restricted Boltzmann Machines (RBMs), we constrain the network connectivity based on a geometric locality principle. Unlike standard ``all-to-all'' RBMs, which entail $\mathcal{O}(N^2)$ connections and are ill-suited for FPGA routing, we employ a spatially local connectivity mask.

\subsection{Connectivity Metric}
\label{sec:supp_4}
We define the connectivity between a neuron $i$ in layer $L$ at lattice coordinate $\mathbf{r}_i$ and a neuron $j$ in layer $L+1$ at $\mathbf{r}_j$ using the Euclidean distance on the periodic lattice (we assign to each hidden and deep layer a 2D geometry isomorphic to the visible lattice, such that every neuron has a defined spatial coordinate):
\begin{equation}
    d(i,j) = \min_{\mathbf{\delta} \in \mathbb{Z}^2} || \mathbf{r}_i - \mathbf{r}_j + \mathbf{L} \cdot \mathbf{\delta} ||_2
\end{equation}
A synaptic connection $W_{ij}$ is non-zero if and only if $d(i,j) \le k$, where $k$ is the tunable connectivity radius. We choose the Euclidean metric because it provides a simple and physically motivated notion of locality on the lattice, enabling sparse and distance-limited connectivity between neurons.

\subsection{Parameter Enumeration}
For the $10 \times 10$ lattice ($N=100$ spins) used in Figure 4:
\begin{itemize}
    \item \textbf{Sparse RBM:} Consists of one visible layer ($N_v=100$) and one hidden layer ($N_h=100$). The total parameters $N_p$ include $N_v + N_h$ biases and the number of active weights defined by $kN_v$.
    \item \textbf{Sparse DBM:} Consists of one visible layer ($N_v=100$), one hidden layer ($N_h=100$), and one deep layer ($N_d=100$). The total parameters $N_p$ include $N_v + N_h + N_d$ biases and two sets of weights, constrained by connectivity radius $k_1$ (between visible-hidden) and $k_2$ (between hidden-deep).
\end{itemize}

Tables~\ref{tab:parameter_counts_rbm}~and~\ref{tab:parameter_counts_dbm} details the exact parameter counts used for the sweep in Figure~4(b,c).

\begin{table}[!ht]
    \centering
    \caption{Detailed parameter breakdown for Sparse RBM architectures evaluated in Figure 4 ($10 \times 10$ Lattice). The connectivity radius $k$ determines the number of active synapses per neuron.}
    \label{tab:parameter_counts_rbm}
    \begin{tabular}{cc|cc}
        \toprule
        \textbf{Radius ($k$)} & \textbf{Neighbors per Neuron} & \textbf{Weights} & \textbf{Total $N_p$} \\
        \midrule
        1 & 5 & 500 & 700  \\
        2 & 13 & 1300 & 1500 \\
        3 & 29 & 2900 & 3100\\
        \bottomrule
    \end{tabular}
    \begin{flushleft}
        \footnotesize{\textit{Note:} ``Neighbors per Neuron" includes the self-connection (same coordinate in adjacent layer) plus spatial neighbors. Total $N_p$ accounts for biases (200 for RBM).}
    \end{flushleft}
\end{table}

\begin{table}[!ht]
    \centering
    \caption{Detailed parameter breakdown for Sparse DBM architectures evaluated in Figure 4 ($10 \times 10$ Lattice). The connectivity radii ($k_1, k_2$) determine the number of active synapses per neuron.}
    \label{tab:parameter_counts_dbm}
    \begin{tabular}{cc|cc|cc}
        \toprule
        \textbf{Radius ($k_1$)} & \textbf{Neighbors per Neuron} & \textbf{Radius ($k_2$)} & \textbf{Neighbors per Neuron} & \textbf{Weights} & \textbf{Total $N_p$} \\
        \midrule
        1 & 5 & 1 & 5 & 1000 & 1300  \\
        2 & 13 & 1 & 5 & 1800 & 2100  \\
        2 & 13 & 2 & 13 & 2600 & 2900 \\
        \bottomrule
    \end{tabular}
    \begin{flushleft}
        \footnotesize{\textit{Note:} ``Neighbors per Neuron" includes the self-connection (same coordinate in adjacent layer) plus spatial neighbors. Total $N_p$ accounts for biases (300 for DBM).}
    \end{flushleft}
\end{table}

\section{Simplified Computational Cost Model for Dual Sampling}
\label{supp_sec:complexity}

We provide a simplified cost model for the dual-sampling algorithm used to estimate local energies and parameter derivatives for DBM neural quantum states in variational Monte Carlo. The goal of this section is to quantify how the wall-clock cost per iteration scales with model size at fixed sampling budgets and sparsity, and to demonstrate how probabilistic computing accelerates the sampling phase. We emphasize that the number of outer samples $N_s$ and inner conditional samples $N_c$ required to reach a prescribed statistical error can, in general, depend on the system size, Hamiltonian properties, sparsity and other parameters. Determining these dependencies requires a separate analysis and is outside the scope of this hardware-focused cost model.

In the following cost analysis, let $N_v$ denote the number of visible spins, $N_h$ the number of hidden units, and $N_d$ the number of deep units, with total number of stochastic units $N_u=N_v+N_h+N_d$. We assume a fixed sparse connectivity with bounded degree $k=O(1)$. Let $N_s$ be the number of outer Monte Carlo samples (visible configurations), and $N_c$ the number of conditional samples drawn in the inner loop for each visible configuration.

For each outer sample, one Gibbs sweep over the visible layer costs $O(N_v k)$. For each of $N_c$ conditional samples, a Gibbs sweep over auxiliary units costs $O((N_h+N_d)k)$ and evaluating all single-spin-flip ratios reusing the same conditional sample costs $O(N_v k)$ (exploiting the locality of the energy function). Thus, the sampling work per outer sample is
\begin{equation}
O\!\left(N_v k + N_c \left[(N_h+N_d)k + N_v k\right]\right),
\end{equation}
and the total sampling work per optimization step is
\begin{equation}
\label{eq:cpu_sampling_cost_revised}
O\!\left(N_s N_c N_u k\right).
\end{equation}
For bounded degree $k=O(1)$, the total work scales as the product of the sampling budget and the system size, $O(N_s N_c N_u)$. Crucially, reusing each conditional sample to evaluate all single-spin-flip ratios avoids the quadratic $O(N_s N_c N_v N_u)$ work (or $O(N_s N_c N_u^2)$ when $N_v = O(N_u)$) that would arise if conditional samples were generated independently for each of the $N_v$ ratios.

On probabilistic hardware (p-computers) implemented on FPGAs, stochastic units can be updated in parallel subject to a fixed update schedule (e.g., graph coloring). Let $C$ denote the number of update phases, which depends on the local connectivity pattern but is independent of $N_u$ for bounded degree graphs. Assuming the full sparse Boltzmann network fits within hardware resources, the wall-clock time for one Gibbs sweep is $O(C)$. Therefore, the sampling \emph{latency} per optimization step scales as
\begin{equation}
\label{eq:fpga_sampling_cost_revised}
O(N_s N_c C).
\end{equation}

Comparing Eq.~(\ref{eq:fpga_sampling_cost_revised}) to Eq.~(\ref{eq:cpu_sampling_cost_revised}), probabilistic hardware removes the explicit $N_u$ factor associated with sequential sampling updates on conventional processors. This is the key distinction between conventional processors and probabilistic hardware: for sparse models with bounded degree and fixed sampling budgets $(N_s, N_c)$, the sampling \emph{latency} on probabilistic hardware scales as $O(N_s N_c C)$, where $C$ depends only on local connectivity and is independent of $N_u$ under the on-chip mapping assumption. Hardware resources (number of p-bits and synapses) scale approximately linearly with $N_u$, but the sampling rate remains constant with system size.

Finally, parameter updates are computed using stochastic reconfiguration (SR) with a matrix-free conjugate gradient (CG) solver. Let $N_p$ be the number of variational parameters and $K_{\mathrm{CG}}$ the number of CG iterations, which may in general vary with $N_u$. Each CG iteration requires sample-averaged matrix-vector products with cost $O(N_s N_p)$, giving
\begin{equation}
\label{eq:sr_cost_revised}
O(K_{\mathrm{CG}} N_s N_p)
\end{equation}
per optimization step. For bounded-degree sparse graphs, $N_p$ scales linearly with system size, i.e., $N_p=O(N_u)$. If translational symmetry is enforced via parameter tying, the effective $N_p$ can be reduced further, but we treat the untied case here.

\end{document}